\documentclass[useAMS,usenatbib]{mn2e}

\usepackage[utf8x]{inputenc}
\usepackage{bm}
\usepackage{latexsym}
\usepackage{tabularx}
\usepackage{dcolumn}
\usepackage{amsmath,amsfonts,amssymb}
\usepackage{graphicx,epsfig}
\usepackage{color}
\usepackage{subfigure}
\usepackage{sidecap}
\usepackage{longtable}
\usepackage{natbib}
\usepackage{float}

\newcommand{\de}{{\rm d}}
\newcommand{\bea}{\begin{eqnarray}}
\newcommand{\eea}{\end{eqnarray}}
\newcommand{\be}{\begin{equation}}
\newcommand{\ee}{\end{equation}}
\newcommand{\f}{\frac}
\newcommand{\df}{\dfrac}
\newcommand{\dl}{\delta}

\newcommand{\bc}{\begin{center}}
\newcommand{\ec}{\end{center}}

\newcommand{\T}{\rule{0pt}{3.6ex}}
\newcommand{\TT}{\rule{0pt}{2.6ex}}

\newcommand{\B}{\rule[-1.0ex]{0pt}{0pt}}

\title{Spatial Clustering of High Redshift Lyman Break Galaxies}
\begin{document}

\author[C. Jose et al.]
{Charles Jose$^1$\thanks{charles@iucaa.ernet.in},
Kandaswamy Subramanian$^1$\thanks{kandu@iucaa.ernet.in},
Raghunathan Srianand$^1$\thanks{anand@iucaa.ernet.in} 
and \newauthor Saumyadip Samui$^2$\thanks{samuis@ukzn.ac.za} \\
$^1$IUCAA,Post Bag 4, Pune University Campus, Ganeshkhind, Pune 411007, India\\
$^2$Astrophysics and Cosmology Research Unit, School of Physics, UKZN, Durban 4001, South Africa}

\maketitle

\begin{abstract}
We present a physically motivated semi-analytic model to understand 
the clustering of high redshift Lyman break galaxies (LBGs). We show 
that the model parameters constrained by the observed luminosity function,  
can be used to 
predict large scale (i.e $\theta\ge 80''$) 
 bias and angular correlation function of galaxies.
These predictions are shown to reproduce the observations
remarkably well.
We then adopt these model parameters to 
calculate the halo occupation distribution (HOD) using
the conditional mass function.
The halo model using this HOD is
shown to provide a reasonably good 
fit to the observed clustering of LBGs 
at both large 
%($\theta>80''$) 
and small ($\theta < 10''$) angular scales
for the whole range of $z=3-5$ and limiting magnitudes.
However, our models underpredict the clustering 
amplitude
at intermediate angular scales, where quasi-linear effects are important.
The average mass of halos contributing to the observed clustering
is found to be $ 6.2 \times 10^{11}$ M$_\odot$ and the characteristic 
mass of a parent halo hosting  satellite galaxies is 
$1.2 \times 10^{12}$ M$_\odot$ for a limiting absolute magnitude of 
$-20.5$ at $z=4$. 
For a given threshold luminosity these masses decrease with 
increasing $z$ and at any given $z$ these are found to increase with 
increasing value of threshold luminosity.
Our physical model for the HOD suggests that
approximately $40\%$ of the halos above a minimum mass $M_{min}$,
can host detectable central galaxies and about 
$5-10\%$ of these halos are likely to also host a detectable satellite.
These satellites form typically a dynamical timescale prior to
the formation of the parent halo.
The small angular scale clustering is mainly due to central-satellite 
pairs rather than few large 
clusters.
It is
quite sensitive to changes in the duration of star formation  
in a halo and hence could provide a probe of this quantity.
The present data favor star formation in a halo lasting typically for a 
few dynamical time-scales,
with 50\% of stars formed in a time 
$T \sim 300-500$ Myr for dark matter halos 
that collapse in the redshift range of $5.5-3.5$.
Our models also reproduce different known trends
between parameters related to star formation.
\end{abstract}

\begin{keywords}
cosmology: theory -- cosmology: large-scale structure of universe -- galaxies: formation -- galaxies: 
high-redshift -- galaxies: luminosity function -- galaxies: statistics -- galaxy: haloes
\end{keywords}

\section{Introduction}
Over the past decade there has been a growing wealth of observations 
probing the properties of high redshift galaxies.
Various surveys, using the Lyman break color selection technique
\citep{madau_96, steidel_96_1,steidel_98,steidel_98_1}, 
have detected a substantial number of  
high redshift galaxies, up to $z\sim 10$. This has resulted in
reasonably good estimates of luminosity functions (LF) of these
Lyman break galaxies (LBG) up to $z\sim 8$ 
\citep{bouwens_07_LF_z46,bouwens_07_LF_z710, reddy_08_LF,vanderburg_10_LF} 
and also LBG clustering up to $z \sim 5$ 
\citep{giavalisco_dickinson_01,porciani_giavalisco_02,ouchi_04_acf,
adelberger_steidel_05,ouchi_hamana_05_acf,kashikawa_06_acf,
lee_giavalisco_06_acf,hildebrandt_09_acf,savoy_11_acf,bielby_11_acf}.
It is important to explain these observations and 
understand their implications for galaxy formation.

In the hierarchical model of structure formation galaxies 
form in virialized dark matter halos. 
These inturn result from the growth and gravitational collapse of initial 
Gaussian density perturbations. Thus the statistical properties of galaxies are
determined by the statistics of the parent halo population, given a model 
for how stars form inside these halos. The properties of dark matter halos 
are quite well understood using N-body simulations \citep{springel_white_05} 
and analytical models like the 
halo model of large scale structure \citep{cooray_sheth_02}. 
These approaches provide 
the abundance, spatial distribution and merger 
history of dark matter halos. Numerical simulations also
suggest a possible universal dark matter halo density profile, 
NFW profile \citep{NFW_97}. 
Given the above inputs on
dark matter halo properties and a specific model of 
galaxy formation inside these halos, it is possible to explain 
the two major observables of galaxies, 
their luminosity function and clustering. 
In addition, such models can throw light on the complex physics of 
galaxy formation, such as rate and duration of star formation, 
feedback mechanisms etc. 

There has been extensive modelling of the luminosity functions and clustering 
of galaxies at low redshifts 
\citep{somerville_99,yang_mo_03, zheng_zehavi_09,zehavi_zheng_11}.
Several studies on understanding the LF of high redshift LBGs have
also been carried out \citep{somerville_01, benson_bower_03,stark_loeb_07,khochfar_silk_07}.
We have been exploring physically motivated semi-analytic models 
%earlier extensively studied simple physical models
of galaxy formation to understand the LFs of LBGs and 
Lyman-$\alpha$ emitters, galactic winds and 
reionization of the intergalactic medium 
\citep{samui_07,samui_08,samui_09,samui_09_lae,samui_10}.
In our model, the luminosity of any galaxy 
is obtained from a physical model of star formation 
rate which depends on the mass and age of the hosting halo.
We then combine this information with the formation 
rate of dark matter halos to obtain the LF of LBGs at various redshifts
(see section 2 for details). 
These models have reproduced the LFs of high $z$ LBGs from
$z=3-7$ reasonably well. 
In addition they constrain the efficiency of star formation and its duration
in LBGs, and have also been used to set tight limits on the neutrino mass 
\citep{charles_11}. 
We now wish to examine if our simple physical
model for galaxy formation, combined with the halo model, can also
explain the clustering of the high $z$ LBGs.

%Recent work on 
%understanding LBG clustering is  
%based on empirical or parametrized models of galaxy formation 
%\citep{bullock_wechsler_02_HOD,hamana_04,hamana_06,conroy_wechsler_06,lee_09}.

Semi-analytical models of clustering involve 
giving a prescription for 
how many galaxies of different luminosities occupy
a dark matter halo of a given mass. 
This is called the Halo occupation distribution (HOD) and is
usually given in a parametrized form 
\citep{jing_98,seljak_00_HOD,scoccimarro_sheth_01,bullock_wechsler_02_HOD,
bullock_02,berlind_02,bosch03,berlind_weinber_03,kravtsov_04,zehavi_weinberg_04,
hamana_04,zehavi_zheng_05,zheng_05,hamana_06,conroy_wechsler_06,lee_09}. 
In our work we calculate the HOD  
without assuming any parametric form.
%using a more physically motivated model.
Combining this
with the dark matter halo abundance, bias and density profile, 
the galaxy clustering can be calculated. 
%The parameters of the model are then varied to fit the
%observed clustering data.

%Our model for star formation inside dark matter halos 
%is that used earlier for understanding the LF of LBGs.
To begin with we assume that each halo can host at most one visible 
central galaxy.
By fitting the observed LF of LBGs,
we find the masses of dark matter halos which host 
an LBG of a given luminosity. 
We then show that our prescription of 
star formation that fits the observed LF of LBGs 
can also simultaneously explain their large scale clustering
 ($\theta \ge 80"$). 
In order to also account for the small angular scale clustering
we calculate how many subhalos
hosting a detectable satellite can form in a bigger
parent halo using the conditional mass function \citep{cooray_sheth_02}
and our star formation prescription.
Thus we provide a method of calculating the HOD 
from first principles, 
%with physically motivated parameters,
which can then be used to predict 
the LBG clustering.
%a physically motivated approach. 
%
%
Using our approach, we show that one can explain both 
the UV LFs and luminosity dependent clustering of LBGs and
gain useful insights into galaxy formation. 

The organization of this paper is as follows. 
In the next section we describe our physical model for computation of the LF of LBGs.  
In Section 3 we focus on the clustering of LBGs on large angular scales.
We then turn to our physically motivated model to calculate the central and satellite 
contributions to the HOD and use these to obtain total angular correlation functions at all angular scales.
Section 5 presents a comprehensive comparison of the total angular correlation function computed 
in various models with observations. A discussion of our results and conclusions are presented 
in the final section. For all calculations we adopt a flat $\Lambda$CDM 
universe with cosmological parameters consistent with 7 year Wilkinson 
Microwave Anisotropy Probe (WMAP7) observations \citep{larson_11_wmap7}. 
Accordingly we assume $\Omega_m=0.27$, $\Omega_\Lambda=0.73$, 
$\Omega_b=0.045$, $h=0.71$, $n_s = 0.963$ and $\sigma_8 = 0.801 h^{-1}$Mpc. 
Here $\Omega_i$ is the background density of any species 
'i' in units of critical density $\rho_{c}$. 
The Hubble constant is $H_0 = 100 h$ km s$^{-1}$ Mpc$^{-1}$

\section{The Star formation rate and Luminosity function} 
\label{section:SFR}

Our aim is to construct a self consistent semi-analytical model, that can 
explain the luminosity functions and clustering of high redshift LBGs. 
In this section, we briefly recall the semi analytical treatment of 
\citet{samui_07} (hereafter SSS07) to model luminosity functions of 
high redshift LBGs [See also ~\citet{samui_09}  (hereafter SSS09); 
\citet{charles_11}] before presenting our prescription to calculate
angular correlation functions at large and small scales. 
In this section we also show the relationship between star formation
rate and stellar mass and stellar mass function predicted by our model.

\begin{figure}
\includegraphics[trim=0cm 0.0cm 0cm 0.0cm, clip=true, width= 8.5cm, height=8.0cm, angle=0]
{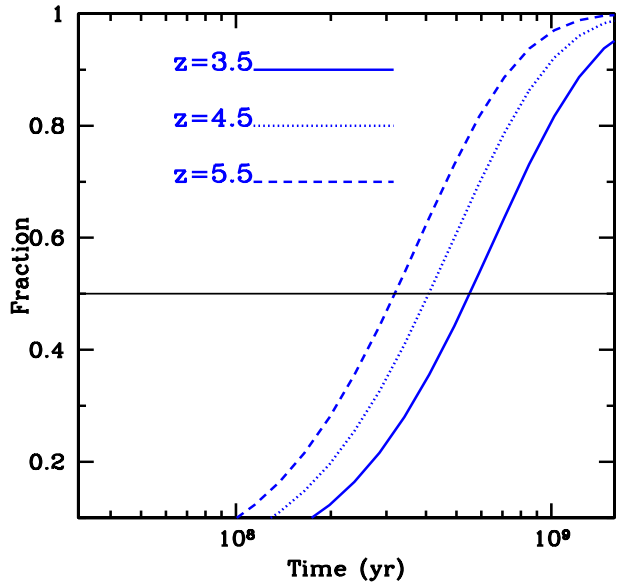} 
\caption{
The fraction of stars formed in a halo in our model
as a function of the age of the halo. 
The three different curves corresponds to halos formed 
at $z=3.5$, $4.5$ and $5.5$, assuming $\kappa=1$.
}
\label{fig:star_frac}
\end{figure}

In the models of SSS07, the star formation rate ($\dot M_{SF}$) in a 
dark matter halo of mass $M$ collapsed at redshift $z_c$ and observed at 
redshift $z$ is given by \citep[see,][]{chiu_00,choudhury_02},
\bea
\dot M_{SF}(M,z,z_c) &=& f_\ast \left(\frac{\Omega_b}{\Omega_m} M \right) 
                \frac{t(z)-t(z_c)}{\kappa^2 t^2_{dyn(z_c)}} \label{eqn:SFR}
 \\ \nonumber
  && \times \exp\left[-\frac{t(z)-t(z_c)}
                {\kappa t_{dyn(z_c)}}\right], 
\eea
where, $f_\ast$ is the fraction of the total baryonic mass that is converted 
into stars over the entire lifetime of the galaxy  
and $t(z)$ is the age of the universe 
at redshift $z$; thus $T(z,z_c)=t(z)-t(z_c)$ is the age of the galaxy at $z$. 
Further, $t_{dyn}(z_c)$  
is the dynamical time scale of a halo collapsing at $z_c$ 
%(Eq.~(3) of SSS07)
and is given by 
\be
t_{dyn}(z_c) = \sqrt{\f{3\pi}{32 G \rho_{vir}(z_c)}},
\label{tdyn}
\ee
where $\rho_{vir}(z_c)= \Delta_c(z_c)\rho_c(z_c)$ with $\Delta_c(z_c)$ being
the over density of the halo at the redshift of collapse, relative to the
critical density $\rho_c(z_c) = [3H^2(z_c)/8\pi G]$.
Typically $t_{dyn}$ at any redshift is about $10\%$ of the Hubble
time at that redshift. Finally, 
$\kappa$ in Eq.~\ref{eqn:SFR}, is a parameter which governs 
the duration of the star formation activity. The star formation rate in a 
halo reaches a peak value when it's age is $\kappa t_{dyn}$. 
Also over the life time of the galaxy a total baryon mass of 
$f_\ast M ( \Omega_b/\Omega_m)$ will be converted into stars in any halo of mass $M$. 
In Fig.~\ref{fig:star_frac}, we show the fraction of the 
stars formed inside a halo (i.e $M_* / [f_\ast M ( \Omega_b/\Omega_m)]$) 
as a function of the age of the halo (here $M_*$ is the stellar mass in
the halo). From the figure one can see that 50 \% of 
the stars are already in place within a time-scale of $T \sim 300-500$ 
Myr inside dark matter halos that collapses in the redshift range of 
$5.5-3.5$ for $\kappa = 1$. While the star formation
can last for few 10$^9$ yrs, the period over which the galaxy is 
detectable depends on the halo mass and the luminosity threshold 
of the observations.

The stars are formed with a Salpeter IMF in the mass range 
$1 - 100 ~M_\odot$. The population synthesis code {\sc Starburst99} 
\citep{starburst_99} is used to obtain the rest frame luminosity  
($l_{1500}$) at 1500 \AA\ as a function of time of a galaxy undergoing 
a burst of star formation. 
The assumed star formation rate of a galaxy, as given in Eq.~(\ref{eqn:SFR}), 
is then convolved with this burst luminosity to get the time evolution of the 
luminosity, $L_{1500}$, of an individual star forming galaxy 
(See Eq.~(6) and Figure 1 of SSS07)
\be
L_{1500}(T) = \int_T^0 \dot M_{SF} (T-\tau) l_{1500}(\tau)d\tau 
\label{eqn:L}.
\ee

\begin{figure*}
\includegraphics[trim=0cm 0.0cm 0cm 8.5cm, clip=true, width =16.5cm, height=7.5cm, angle=0]
{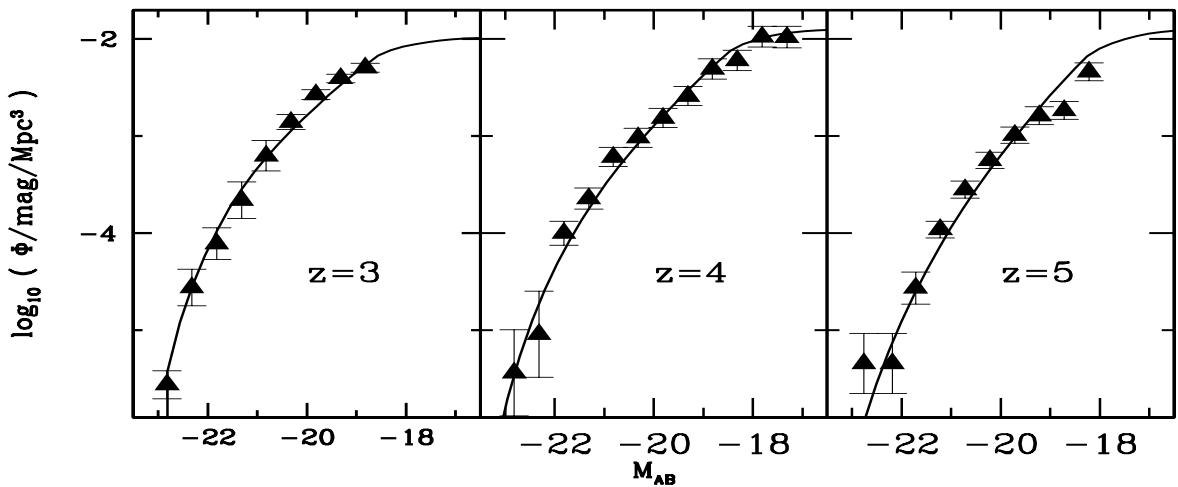} 
\caption{Comparison of observed UV LF of LBGs at three different redshifts 
with our best fitted model predictions. The observed data points and error 
bars are from \protect \cite{reddy_08_LF} (for z=3) and  \protect 
\cite{bouwens_07_LF_z46} (for z=4 \& 5).
}
\label{fig:lfz}
\end{figure*}

Due to dust absorption, only a fraction ($1/\eta$) of $L_{1500}$ produced 
by the stars manages to escape the galaxy.
This luminosity ($L = L_{1500}/\eta$) is then converted to a standard 
absolute AB magnitude $M_{AB}$,  using the equation given by \citep{oke_83},
to enable direct comparison with the observed data. 
Having obtained the $M_{AB}$ of individual galaxies we can compute
the luminosity function $\Phi(M_{AB}, z )$ at any redshift $z$ 
using,
\bea
\Phi(M_{AB}, z ) dM_{AB}&=& \int\limits_z^\infty dz_c  \frac{dn(M(M_{AB}),z_c)}{dz_c}
\frac{dM}{dL_{1500}} \\ \nonumber  &&\times~ \frac{dL_{1500}}{dM_{AB}} ~dM_{AB}.
\eea
Here $\de n(M,z_c)/\de z_c =  \dot n(M,z_c) dt/dz_c$, and $\dot n(M,z_c) dM$ is 
the formation rate of halos in the mass range $(M, M+dM)$ at redshift $z_c$. 
SSS09 modelled this formation rate as the time derivative of 
\citet{sheth_tormen_99} (hereafter ST) mass function as they are found to be 
good in reproducing the observed LF of high-$z$ LBGs. Therefore we use 
$\dot n(M,z_c) = \de n_{ST}(M,z_c)/\de t$  where 
$ n_{ST}(M,z_c)$ is the ST mass function at $z_c$. 
Also note that we use the notation $n(M)$ for $dn/dM$ for convenience.

Star formation in a given halo also depends on the cooling efficiency of 
the gas and various other feedback processes. We assume that gas in halos with 
virial temperatures ($T_{vir}$) in excess of $10^4$ K can cool 
(due to recombination line cooling from hydrogen and helium) and collapse 
to form stars. However the ionization of the IGM by UV photons increases 
the temperature of the gas thereby increasing the Jean's mass for collapse. 
Thus in ionized regions, we incorporate this feedback by a complete 
suppression of galaxy formation in halos with circular velocity 
$v_c \leqslant 35$ km s$^{-1}$ and no suppression with 
$v_c  \geqslant V_u = 95$ km s$^{-1}$ 
\citep{bromm_02}. For intermediate circular velocities, a linear fit from 
$1$ to $0$ is adopted as the suppression factor [\cite{bromm_02}; 
see also \cite{benson_02,dijkstra_04}, SSS07]. 
SSS07 found that this feedback mechanism naturally leads to the 
observed flattening of the LF at the low luminosity end.
 
In our models, we also incorporate the possible Active Galactic Nuclei (AGN) 
feedback that suppresses star formation in the high mass halos, by multiplying 
the star formation rate by a factor $[1+M/M_{agn}]^{-\beta}$. This 
decreases the star formation activity in high mass halos above a 
characteristic mass scale $M_{agn}$, 
which is believed to be $\sim 10^{12} M_\odot$ 
\citep[see][]{bower_06,best_06}. 
In our models we consider $\beta = 0.5$ 
and  $M_{agn}$ as a free parameter.
Note, SSS07 used AGN feedback with $\beta = 3$, but
suppressed the halo formation rate instead of star formation rate
as we do here.

A crucial parameter of our model is $f_\ast/\eta$ which governs the mass to 
light ratio of the galaxies at any given redshift. 
A number of recent works have tried to measure the amount 
of dust obscuration of UV luminosity (parametrized here 
by $\eta$) at high redshifts by fitting the spectral energy 
distribution (SED) of LBGs \citep{reddy_pettini_12,wilkins_gonzalez_12,
gonzalez_bouwens_12,bouwens_illingworth_12}.
They suggest a tentative evidence for the dust correction to
increase with luminosity at a given redshift. 
However introducing such a trend, which is not yet well established
(at least quantitatively),
would add another source of uncertainty in our models.
Therefore, for simplicity, in our models
we assume $\eta$ to be luminosity independent
at any given redshift.
It is also been suggested
that the dust corrections may evolve with redshift.
We do take into account this average evolution of $\eta$ with $z$
as given by 
\citet{stark_ellis_09_z46,gonzalez_bouwens_12,reddy_pettini_12}. 
Note that for a luminosity independent $\eta$, we only need the 
combined parameter $f_\ast/\eta$ to fit the LFs of LBGs.
Moreover, as we will show below, 
the clustering predictions are also determined solely  
by the combined parameter $f_\ast/\eta$. 
%and hence we will not try to constrain $f^\ast$ independently in this paper.
We describe the free parameters of our model in Table. \ref{tab0}. 
\begin{table}
\caption{ 
The free parameters of our physical model.}
\begin{tabular}{cl}\hline

\hline
Parameter    &Description \\
\hline   
$f_\ast/\eta$ &Related to the light to mass ratio. This\\ 
~             &parameter is assumed to be independent \\
~              &of the mass of the halo.\\
$M_{agn}$     &Determines the mass scale of AGN feedback.\\
~             &The AGN feedback is assumed to suppress\\
~             &the star formation in dark matter halos by a \\
~	      &factor $(1+M/M_{agn})^{0.5}$.  \\
$\kappa$      &Determines the typical duration of star in \\
~             &individual dark matter halos. The star formation\\
~	      &rate in a halo reaches a peak value when it's \\
~	      &age is $\kappa t_{dyn}$.  \\
$\Delta t_0$  &This parameter, defined in Section~4, is used \\
              & for calculating small angular scale clustering.\\
              &$\Delta t_0$ is the minimum time difference\\
              & between the formation epochs of a parent \\
              & halo and sub-halos hosted by it.\\
%              &Our model assumes that if $t(z_p)$ is the age of the\\  
%~             &universe when a parent halo is collapsed and $t(z_s)$\\
%~	      &is the age of the universe when a subhalo is formed \\ 
%~	      &inside the same parent halo then $t(z_p)-t(z_s) \geq \Delta t_0$ \\
\hline  
\end{tabular}
\label{tab0}
\end{table}

The parameter $f_\ast/\eta$ at each redshift
is fixed by fitting the observed luminosity 
function of LBGs using $\chi^2$ minimization. 
In this way our physically motivated model for star formation 
gives the relationship between 
the halo mass (M) and the luminosity (L) of the galaxy it hosts.
It is important to note that 
our prescription of star formation naturally introduces a scatter in 
the  M-L relationship because halos of mass $M$ forming at different 
redshifts, produce different luminosities at the redshift of observation. 
SSS09 showed that this M-L relationship can successfully explain 
the luminosity functions of high redshift LBGs.

In Fig.\ref{fig:lfz}, we show the observed luminosity function of 
LBGs together with our best fitted model results for three different redshifts. 
As a fiducial model we have chosen 
$\kappa = 1.0$ at all redshifts and three different values for 
$M_{agn}$, $0.8\times 10^{12}M_\odot$, $1.5\times 10^{12}M_\odot$, and 
$3.0\times 10^{12}M_\odot$ at redshifts 3, 4 and 5 respectively. 
With these parameters the observed luminosity functions 
is well reproduced for $f_\ast/\eta$ of 0.042, 0.038 
and 0.032 respectively for $z=3,4$ and $5$.
The $\chi^2$ corresponding to the best fit luminosity function 
at redshifts $3,~4$ and $5$ are $10.80,~8.58$ and $10.51$ 
(with corresponding reduced $\chi^2$ of $1.35,~0.78$ and $1.31$) 
respectively. 
These values of $f_\ast/\eta$ are used below when we map the
clustering of dark matter halos to that of galaxies.

%In our model, the star formation rate (SFR) in a galaxy 
%is solely determined by the mass and age of the hosting 
%dark matter halos. 
Recent advances in multi-band deep field 
observations allow one to estimate SFR and stellar mass ($M_*$) 
in individual LBGs and their global stellar mass function using 
SED fitting. These are used to establish trends between
SFR and $M_*$ \citep{stark_ellis_09_z46,gonzalez_labbe_11,
reddy_pettini_12, wilkins_gonzalez_12, gonzalez_bouwens_12, 
bouwens_illingworth_12}.
Note the individual values  of these derived
quantities need not be accurate as they depend on  the models
used to generate the SEDs. However, the observed trends may 
depend weakly on the SED model parameters. 
%
%
%There has been a growing number of other observables related to
%high redshift galaxy formation and evolution, like
%the SFR-stellar mass ($M_*$) correlation, its evolution and the stellar
%mass function. 
Although the focus of our paper is on LBG clustering, 
it would be of interest to compute these quantities and trends
discussed above.
%However, the derived values of these quantites depend on the input 
%parameters used to generate SEDs.
%
%these are derived quantities
%
%
%erived their absolute
%values depend on the input parameters used to generate SEDs.
%Therefore, we mainly check whether our models consistently
%reproduce the trends found by the SED fitting methods (REF).
%
%We will give the details of such a study in a future work, but give here
%a few of the preliminary results.
%It is possible to test this physical model by comparing it with many  
%observables related to high redshift galaxy evolution. 

%For example, i
If we use the average value of $\eta \sim 4,~2.2,~2$ 
at $z=3,~4,~5$ estimated from \citet{bouwens_illingworth_12,reddy_pettini_12}, we get
$f_\ast \sim 0.17,0.08,0.06$ at these redshifts.
This implies an increase in the fraction of baryons converted to 
stars with time. 
Using Fig.~\ref{fig:star_frac}, we can infer that about $3-8\%$ 
of the total baryons 
in a typical galaxy is converted to stars
over a timescale of $300-500$ Myr for $z=5-3$.

Using the derived values of $f_\ast$ we
can calculate $L_{UV}$-stellar mass ($M_*$) relation from our model
at any $z$.
%
%The results are shown in Fig~\ref{fig:mstar-sfr} for $z$ = 4.
%For these calculations we consider only galaxies in our models
%that are having absolute magnitude,$-18\ge M_{AB}\ge -23$.
Our models capture a clear correlation between the 
two quantities as found by the previous authors 
%\citep{gonzalez_bouwens_2012}.
\citep{stark_ellis_09_z46,labbe_gonzalez_10,gonzalez_labbe_10,gonzalez_labbe_11,
gonzalez_bouwens_12,lee_ferguson_12,reddy_pettini_12}
For example, at $z=4$, we find a mean trend which can be approximated 
with a powerlaw $M_* \propto L_{1500}^{1.5}$.
Also at $M_{AB} = -20$, we find $M_*\sim 10^9 M_\odot$.
These compare reasonably with the observations presented by
\cite{gonzalez_labbe_11} (Figure 1), 
who find a relation $M_* \propto L_{1500}^{1.7}$,
with also $M_* \sim 10^9 M_\odot$ at $M_{AB} = -20$.
We can also compute the global stellar mass function 
for the galaxies in any luminosity range.
For galaxies 
with absolute magnitude in the range $-18\ge M_{AB}\ge -23$, 
at $z=4$ and for $M_*\ge 10^9$ M$_\odot$, we find that the stellar mass 
function can be
approximated by a powerlaw with a slope of $\sim$ $-1.5$. We also find
the abundance of galaxies per dex in $M_*$ at $10^9 M_\odot$ is 
about $10^{-2.2}$ per Mpc$^3$.
This compares reasonably well with the results of \cite{gonzalez_labbe_11}
who find a slope of the stellar mass function $-1.4$ to $-1.6$ and 
galaxy abundance $10^{-2.5}$ per Mpc$^3$ per dex.
We defer a detailed discussion of these issues to a future work
as our focus here is on the spatial clustering of LBGs. 

Thus our 
physically motivated models capture
the basic trend seen based on SED fitting analysis.
In passing we mention that the specific star formation
rate (SSFR) calculated in our models and its evolution with
redshifts are found to be consistent with the trends 
quoted in the literature \citep{gonzalez_labbe_11,
gonzalez_bouwens_12, bouwens_illingworth_12}.
As our main focus of this
work is to understand the spatial clustering of LBGs
at different redshifts we differ a detailed discussions
on SFR, $M_*$, SSFR and their redshift dependence to
a future work.
%
%
%
%compares reasonably well with that obtained by \citet{gonzalez_bouwens_12} 
%(see Fig. 3 of their paper). Similarly  we have shown in Fig~\ref{fig:smf} 
%the stellar mass function (SMF) at $z=4$, 
%obtained from our model, where we have included only those LBGs 
%with $-22 < M_{AB}< -18$. (The cut-off at faint magnitudes induces 
%a cut-off at the lower mass end of the SMF). 
%This SMF again compares reasonably well with the observed data 
%of \citet{gonzalez_labbe_2011_smf}. 
%In particular the slope of the stellar mass function from our models,
%for $M_* > 10^9 M_\odot$, is 
%$\sim$ $-1.5$ which is consistent with the results of 
%\cite{gonzalez_labbe_2011_smf}.
%We also find that redshift evolution of the specific 
%star formation rate from $3<z<7$ is consistent with that
%shown in \cite{reddy_pettini_2012,bouwens..}. 
%The details of our models will be given in a separate work.
%\begin{figure}
%\includegraphics[trim=0cm 0cm 0cm 0cm, clip=true, width =8.5cm, height=7cm, angle=0]
%{avg_sfrz4.ps} 
%\caption{
%SFR - stellar mass ($M_*$) relation predicted by our physically motivated 
%models at $z=4$. The solid line represents the mean SFR of galaxies 
%of a given stellar mass and 
%the dotted lines show the $1\sigma$ (68.4 \%) scatter from this mean value.
%}
%\label{fig:mstar-sfr}
%\end{figure}
%\begin{figure}
%\includegraphics[trim=0cm 0cm 0cm 0cm, clip=true, width =8.5cm, height=7cm, angle=0]
%{SMFz4.ps} 
%\caption{
%The stellar mass function at $z=4$ as predicted by our models.  
%}
%\label{fig:smf}
%\end{figure}

\section{Clustering of LBGs at large angular scales}
\label{sec:xi_linear}
We couple the semi-analytic models described above
with the halo model to compute the correlation function of high-$z$ LBGs.
In order to calculate the galaxy-galaxy correlation function on all scales 
one requires a full knowledge of the halo occupation distribution (HOD), 
which describes the conditional probability $P(N|M)$ for $N$ galaxies of 
a given type to reside inside a halo of mass $M$ 
\citep{bullock_02,berlind_02}. On scales much bigger than the virial 
radius of a typical halo, 
the clustering amplitude is dominated by 
correlation between galaxies inside separate halos. On the other hand, 
on scales smaller than the typical virial radius of a dark matter halo, 
the major contribution to galaxy clustering is from galaxies residing 
in the same halo. These separate contributions to two point correlation 
function are called 2-halo and 1-halo terms respectively. In this section 
we concentrate on the 2-halo term.
 
The first moment of halo occupation is the mean number of galaxies of a 
given type inside a parent halo. It has contributions from both
central and satellite galaxies \citep{zheng_05,zehavi_zheng_11}. In the 
calculation of LF in the previous section we have assumed that 
each halo hosts a single star forming galaxy. 
However, the detectability 
of this galaxy depends on its age, the limiting luminosity of 
the observations and 
feedback processes introduced in the previous section. 
Neglecting the one halo term is a 
good approximation for correlation functions on large scales because 
(i) on largest scales the clustering is insensitive to the galaxy 
distribution inside a halo and is dominated by the two halo term,  
(ii) using our model, we later show that, for the LBGs we consider, 
the total number of satellite galaxies of a 
particular luminosity is much less than the total number of central galaxies 
of the same luminosity in a large volume.  Thus on large scales the contribution to clustering 
due to satellite galaxies is not significant compared to the clustering of 
central galaxies. 
  
The galaxy power spectrum on large scales (small $k$) due to the
two halo term alone, assuming 
linear bias, is given by \citep{cooray_sheth_02}
\be
P^{2h}_g(k,z) = b_g^2(k,z) P_{lin}(k,z), 
\label{eqn:pk}
\ee
where $P_{lin}(k,z)$ is the linear dark matter power spectrum. The scale 
dependent, galaxy number weighted, halo bias or in short galaxy bias,
$b_g(k,z)$, is given by 
\be
b_g(k,z) = \f{1}{n_g(z)}\int b(M,z)n(M,z) u(k,M,z) dM.
   \label{eqn:bias}
\ee
As before, $n(M,z)$ is the ST halo mass function, $n_g(z) = \int n(M,z) dM$ 
is the number density of galaxies and $b(M,z)$ is the mass dependent 
halo bias factor provided by the fitting function of Sheth and Tormen 
\citep{sheth_tormen_99,cooray_sheth_02}. It has the following 
functional form 
\be
b(M,z) = 1+ \df{q\nu(M,z)-1}{\dl_c(z)} + \df{2p/\dl_c(z)}{1+(q\nu(M,z))^p},
\label{eqn:halo_bias}
\ee
where, $\nu(M,z) = \dl_c(z)/\sigma(M)$, $\sigma(M)$ is the linearly 
extrapolated rms density fluctuation on any mass scale $M$ 
and $\dl_c(z)$ is the critical density required for collapse 
at $z$. Here $\dl_c(z) = D(z) \dl_c(z=0)$ with $\dl_c(z=0) = 1.686$, 
where $D(z)$ is the linear growth factor in a $\Lambda$CDM universe. 
Also we use $p=0.3$ and $q=0.707$ as given by \citet{sheth_tormen_99}.
%Sheth and Tormen \citep{cooray_sheth_02}. 
In Eq.~(\ref{eqn:bias}) $u(k,M)$ is the Fourier 
transform of dark matter density profile normalized by it's mass, 
i.e $u(k,M,z) = \bar \rho(k,M,z)/M$. 
We assume the Navarro, Frenk and White (NFW) 
profile \citep{NFW_97,bullock_kolatt_01} for the dark matter density 
distribution in halos.  
On scales much greater than the virial radius of a 
typical halo of mass $M$, $u(k, M,z) = 1$. Thus galaxy bias, $b_g(k,z)$, 
on large scales (much larger than the virial radius of typical 
collapsed halos) or small $k$, will be independent of $k$.

The actual clustering observations are for a galaxy sample with 
an apparent magnitude below some threshold (or a corresponding  
lower luminosity threshold $L_{th}$). 
So $b_g(k,z)$ in Eq.~(\ref{eqn:bias}) should be calculated for 
only galaxies which have luminosity greater than $L_{th}$. 
We use our model described in Section~\ref{section:SFR} for calculating the 
time dependent luminosity of
a galaxy hosted by a dark matter halo of mass $M$. This can then be used  
to obtain the galaxy bias for LBGs with the luminosity threshold $L_{th}$, 
at any redshift of observation $z$ as, 
\bea
b_g(k,L_{th},z) &=& \f{1}{n_g(L_{th},z)}\int^\infty_0 dM b(M,z)u(k,M,z) \times \nonumber \\ 
          &&\int^{\infty}_z\ dz_c \Theta\left(L(M,z_c,z) - L_{th}\right) \f{dn(M,z_c)}{dz_c}. \nonumber \\ 
   \label{eqn:biasL}
\eea
Here, $L(M,z_c,z)$ is the luminosity  of a galaxy of mass $M$ at $z$, 
formed at $z_c$. The theta function, $\Theta\left(L(M,z_c,z) - L_{th}\right)$,
in the above equation ensures that the galaxy bias will have contribution 
only from those galaxies formed at $z_c$ and that shine above 
the threshold luminosity $L_{th}$ at $z$. 
If we do not have this constraint imposed by the $\Theta$ function, the integral over $z_c$ is just 
$\int^{\infty}_z\ dz_c dn(M,z_c)/dz_c = n(M,z)$ and the Eq.~(\ref{eqn:biasL})
reduces back to Eq.~(\ref{eqn:bias}).
Also $n_g(z,L_{th})$ is the number density of galaxies having luminosity
in excess of the limiting luminosity, $L_{th}$, and is given by,
\bea
n_g(L_{th},z) = \int_0^\infty dM \int^{\infty}_z\ dz_c \Theta\left(L(M,z_c,z) - L_{th}\right) \nonumber \\ 
 \times \f{dn(M,z_c)}{dz_c}.     
\eea

\begin{figure}
\includegraphics[trim=0cm 0cm 0cm 0cm, clip=true, width =8.0cm, height=7.5cm, angle=0]
{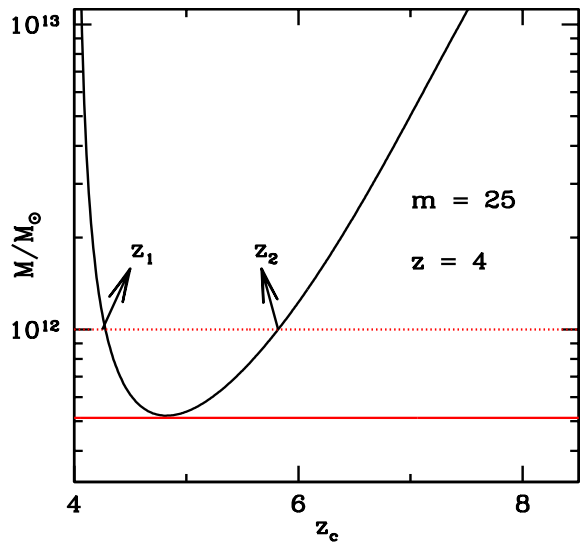} 
\caption{Masses of the dark matter halos that shine with an apparent magnitude 
$m=25$ (or absolute magnitude $M_{AB}=-21.1$) 
at $z=4$ as a function of their formation (collapse) redshift.
We can clearly see that halos of a particular mass formed at two 
different redshifts can shine with the same brightness at $z=4$. 
We refer these two redshifts as  $z_1$ and $z_2$. 
For example, when $M=10^{12} M_\odot$ 
(shown by dotted red horizontal line )
these two redshifts are approximately 4.27 and 5.82. 
The figure also shows the minimum mass 
of the galaxy that can produce an apparent magnitude 25 at $z=4$ 
(in red horizontal solid line).
This mass is roughly $5.2 \times 10^{11}  M_\odot$ and 
is formed at $z_c \sim 4.8$.
}
\label{fig:zc_mass}
\end{figure}

In our prescription of star formation, 
at any given redshift $z$, a galaxy hosted by a halo of mass $M$ will
shine with a given luminosity at two different ages, 
when its star formation is either
in the rising phase or in the declining phase.
As a result there are two redshifts of formation 
$z_1(L_{th},M,z)$ and $z_2(L_{th},M,z)$, 
such that the galaxy will produce an observed luminosity $L_{th}$ at $z$. 
We demonstrate this in Fig. \ref{fig:zc_mass},
where we have plotted the halo mass $M$ that can host a galaxy of luminosity 
$L_{th}$ (corresponding to an apparent magnitude $m = 25$) at $z=4$ 
against the redshift of formation of the halo. 
The figure clearly shows that halos of any mass (above a minimum
mass $ M_{min}$) formed at two different 
redshifts can shine with the same luminosity at $z=4$.
For example, when $M=10^{12} M_\odot$ these two redshifts are approximately 
4.27 and 5.82 (shown by the abscissa of intersection of the horizontal red dotted 
line with the curve). A halo of mass $10^{12} M_\odot$ formed between these 
redshifts will shine with a magnitude brighter than $M_{AB}= -20.1$ at $z=4$. 
For the case illustrated above, this minimum mass is roughly 
$5.2 \times 10^{11}  M_\odot$ and this halo has to be formed at $z_c \sim 4.8$ 
for it to have a luminosity $L_{th}$ at $z=4$.  In our models, the exact
values of $z_1$, $z_2$ and  $ M_{min}$ for a given observed luminosity will
depend only on $f_*/\eta$ that we fix by fitting the luminosity function
as discussed before.

Our model of star formation ensures that halos of mass $M> M_{min}$ collapsing 
between $z_1(L_{th},M,z)$ and $z_2(L_{th},M,z)$ will always shine brighter than 
$L_{th}$ at $z$, while galaxies formed outside these intervals do not.
Thus the $\Theta$ function in Eq.~\ref{eqn:biasL}, is unity for $M> M_{min}$ and
$z_1 \le z_c \le z_2$ and zero otherwise. Therefore $b_g(k,L_{th},z)$ in Eq.~(\ref{eqn:biasL})
can be written as, 
\bea
b_g(k,L_{th},z)  &=& \f{1}{n_g(L_{th},z)}\int^\infty_{M_{min}} dM b(M,z)u(k,M,z) \nonumber \\
&& ~~\int^{z_1}_{z_2}dz_c \f{dn(M,z_c)}{dz_c}. 
   \label{eqn:biasL2}
\eea
At this point, given $M_{min}$, $z_1$ and $z_2$ one can compute $b_g(k,L_{th},z)$. 
As these parameters can be fixed by the parameters governing the star 
formation in a halo, the observed luminosity function alone in principle will allow 
us to predict $b_g(k,L_{th},z)$ uniquely.

We present our results for the fiducial set of model parameters 
that reproduces the observed luminosity function.
In Table~\ref{tab1} we present the predicted asymptotic ($k\longrightarrow 0$) 
value of  $b_g(L_{th},z)$
at three different redshifts (and three luminosity ranges each) using  $f_\ast/\eta$ 
values obtained by fitting the observed luminosity function of LBGs at 
these redshifts. 
This table also gives apparent magnitude cut-off (m), corresponding absolute 
magnitude cut-off ($M_{AB}$), $M_{min}$ and $n_g(L_{th},z)$ as given in Eq.~8.

\begin{table}
\caption{
Asymptotic values of $b_g(L_{th},z)$ predicted uniquely by 
using model parameters that best fit the high $z$ LFs.
These parameters are
(i)$f_\ast/\eta$, an indicator of the light to mass ratio at any redshift
(ii) $M_{min}(L_{th},z)$, the minimum mass of a galaxy that can shine brighter 
than a given luminosity threshold $L_{th}$ at redshift of observation, 
(iii) $n_g(z,L_{th})$, the number density of galaxies with luminosity 
greater than $L_{th}$ in units of $ 10^{-4} (h/Mpc)^3$ and
$b_g(L_{th},z)$, the luminosity dependent galaxy bias  
are tabulated for three limiting apparent magnitudes at each redshift. 
The last column is the galaxy bias given by \protect \cite{hildebrandt_09_acf}, 
after correcting for the larger $\sigma_8$ adopted by them.
}
%\label{table1}
\begin{tabular}{cccccccc}\hline
$z$ &$f_\ast/\eta$ &$m$  &$M_{AB}$\T \B\B \B\B &$M_{min}/M_\odot$   &$n_g$ &$b_g$  &$b_g(H)$  \\

\hline  
\T~   &~      &24.5  &-21.1      &$7.6\times10^{11}$   & $5.81$  &3.63 &4.50 \\
  3   &0.042  &25.0  &-20.6      &$4.2\times10^{11}$   & $14.2$  &3.23 &3.22 \\
\B~   &~      &25.5  &-20.1      &$2.5\times10^{11}$   & $30.3$  &2.92 &2.67 \\
\T~   &~      &25.0  &-21.1      &$5.2\times10^{11}$   & $3.66$  &4.69 &5.14 \\
  4   &0.038  &25.5  &-20.6      &$3.1\times10^{11}$   & $9.17$  &4.22 &4.25 \\
\B~   &~      &26.0  &-20.1      &$2.0\times10^{11}$   & $20.8$  &3.82 &3.51 \\
\T~   &~      &25.5  &-21.0      &$4.0\times10^{11}$   & $1.56$  &5.95 &7.81 \\
  5   &0.032  &26.0  &-20.5      &$2.5\times10^{11}$   & $4.32$  &5.36 &6.08 \\
\B~   &~      &26.5  &-20.0      &$1.5\times10^{11}$   & $10.8$  &4.90 &5.09 \\

\hline
\end{tabular}
\label{tab1}
\end{table}

We compute the luminosity dependent galaxy power spectrum, 
$P_g^{2h}(k,L_{th},z)$, by substituting $b_g(k,L_{th},z)$ from 
Eq.~(\ref{eqn:biasL2}) into Eq.~(\ref{eqn:pk}). We then have 
\be
P_g^{2h}(k,L_{th},z) = b^2_g(k,L_{th},z) P_{lin}(k,z).
\label{eqn:pgl}
\ee
The corresponding luminosity dependent two point correlation function 
of galaxies with luminosity greater than $L_{th}$ at $z$, can now
be calculated using \citep{peebles_80}
\be
\xi^{2h}_g(r,z,L_{th}) = \int_0^\infty \f{dk}{2\pi^2} 
k^2 \f{\sin(kr)}{kr} P^{2h}_g(k,z,L_{th}).     
\label{eqn:xir-2h}
\ee
We compute luminosity dependent angular correlation function 
$w(\theta,z)$ from the spatial correlation function using Limber 
equation \citep{peebles_80}
\be
w(\theta,z) = \int_0^\infty dz'~ N(z') \int_0^{\infty} dz''~N(z'') 
\xi_g\left(z,r(\theta; z',z'')\right) 
\label{eqn:limber}
\ee
where $r(\theta; z',z'')$ is the comoving separation between two points at $z'$ and $z''$ 
subtending an angle $\theta$ with respect to an observer today. Here we have also incorporated 
the normalized redshift selection function, $N(z)$, 
of the observed population of galaxies. In Eq.~(\ref{eqn:limber}) we neglect 
the redshift evolution of clustering of the galaxies detected around $z$.
Hence the spatial two-point correlation function $\xi_g(r,z)$ is 
always evaluated at the observed redshift.

\begin{figure*}
\begin{center}
\includegraphics[trim=0cm 0cm 0cm 0cm, clip=true, width =17cm, height=15cm, angle=0]
{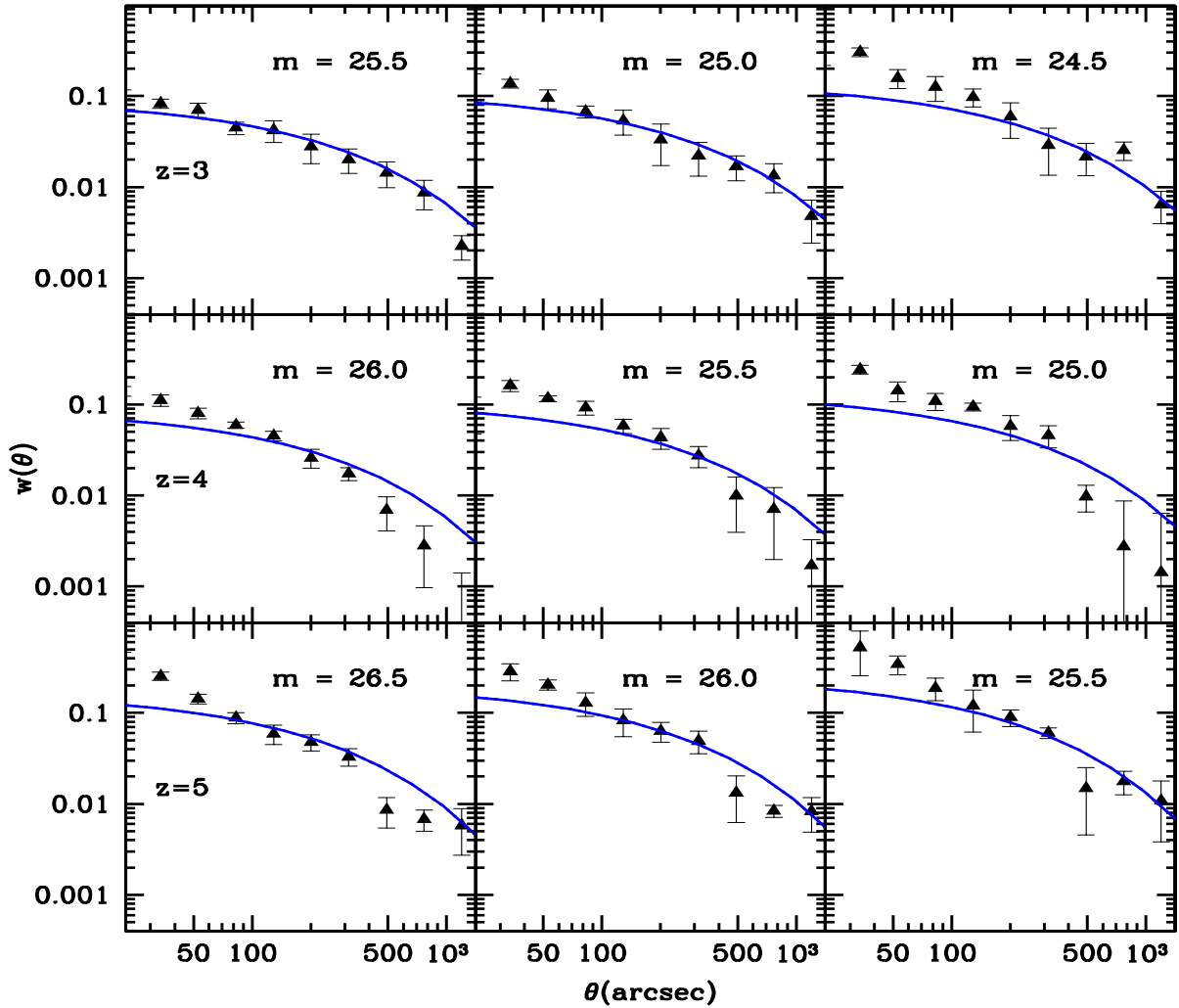} 
\caption{The large scale angular correlation function of LBGs at redshifts 3, 4 
and 5 for various limiting magnitudes. 
Each row corresponds to a particular redshift, which is labelled in the first panel
of that row. In each row there are three panels showing the clustering 
predictions for 
three limiting galaxy magnitudes that are labelled in each panel.
The blue solid curves are our model predictions of galaxy angular correlation functions. 
Also shown in blue dotted lines are the angular correlation functions of dark matter density. 
The data points and error bars shown by black triangles 
are from \protect \cite{hildebrandt_09_acf}.
% and those shown by red squares are from \protect 
%\cite{ouchi_hamana_05_acf}.
}
\label{fig:acfz_lin}
\end{center}
\end{figure*}

In Fig.~\ref{fig:acfz_lin} we over plot the angular correlation function
computed in this way using the predicted value of $b_g$ 
from our models that produce best fits to the LF, on the observed angular two 
point correlation functions given by ~\citet{hildebrandt_09_acf}. 
We have also plotted the angular 
correlation function of dark matter density (obtained by putting 
$b_g(k,L_{th},z) =1$) in dotted-blue curves. To compute our correlation 
function we used the  $N(z)$ 
from ~\citet{hildebrandt_09_acf}, which are kindly 
provided by the authors 
($BC_{sim}$ redshift distribution; see Table 4 and Figure 5 of the paper). 
It is clear that at large angular scales 
(i.e $\theta \ge 80''$), where the linear approximation used here 
is valid, our model predictions match well with the observations at
different redshifts and different luminosity (or apparent magnitude)
thresholds.

The large scale galaxy bias is determined by \cite{hildebrandt_09_acf} 
from their data by making power law fits to observed galaxy 
correlation functions, and comparing the corresponding galaxy
variance at $8 h^{-1}$ Mpc ($\sigma_{8,g}$) with $\sigma_8$
computed from the dark matter power spectrum.
In the last column (column 7) of 
Table.~\ref{tab1} we show the 
large scale bias $b_g(H)$ of LBGs obtained this way 
by \cite{hildebrandt_09_acf},
after correcting for the larger $\sigma_8$ adopted by them.
The large scale bias $b_g$ that we predict from our models
agrees well with $b_g(H)$ estimated by \cite{hildebrandt_09_acf}
from their power law fit, for all but the brightest of the LBG samples.
For the brightest sample of LBGs at all redshifts the bias we
predict is systematically lower. Nevertheless, 
%These numbers roughly agree with the values we find here. 
in agreement with \cite{hildebrandt_09_acf}, we also find 
that (i) at a given $z$ the bias increases with 
increasing luminosity and (ii) for a given $L_{th}$ the bias 
increases with increasing $z$. 
However at any given $z$ the spread in $b_g$ 
as a function of $L_{th}$ is less in our case 
compared to that of \cite{hildebrandt_09_acf}. 

We also see from Fig.~\ref{fig:acfz_lin} that the two halo term is not able to
account for the strength of clustering seen on small angular scales 
given by  $\theta < 50''$ at redshift 3 and $\theta < 80''$ at redshifts 4 and 5.
However we have not yet included the 1-halo term in the correlation function, 
arising from correlation between galaxies within any given halo.
We will address this issue in the Section~\ref{sec:xi_total}.   

\subsection{Sensitivity of large angular scale correlation function to 
change in parameters}
\label{sec:2halo_sensitivity}
Here we present the sensitivity of the predicted large  scale  
correlation functions to changes in astrophysical and cosmological parameters. 
To do this, we considered models, where	a single astrophysical or cosmological parameter 
is assigned values around it's fiducial value, keeping all other parameters 
fixed to their fiducial value. 

\subsubsection{Sensitivity to Astrophysics}
\begin{figure}
\includegraphics[trim=0cm 0cm 0cm 0cm, clip=true, width =8.0cm, height=7.5cm, angle=0]
{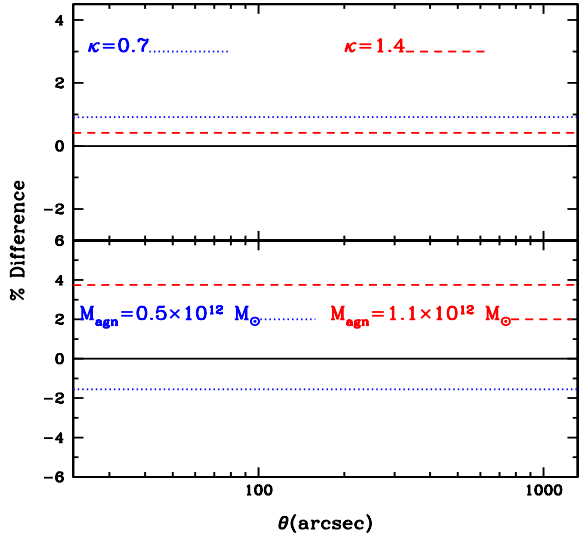} 
\caption{
The percentage change is angular correlation function when one vary 
$\kappa$ (top panel) and $M_{agn}$ (bottom panel) 
around the fiducial model at $z=3$. Here the plots are 
for threshold apparent magnitude 25. For each value of $\kappa$ and 
$M_{agn}$ we use the appropriate $f_\ast/\eta$ values that best fit 
the observed luminosity function. 
}
\label{fig:acf_z4_ratio_astro_2h}
\end{figure}

We begin by showing in Fig.~\ref{fig:acf_z4_ratio_astro_2h} sensitivity of large 
scale clustering predictions to the assumed values of, 
$\kappa$ which determines duration of star formation in a halo 
and $M_{agn}$, a characteristic mass scale corresponding to the AGN feedback. 
Note that, in each case the $f_\ast/\eta$ gets 
automatically fixed when we 
fit the observed luminosity function. The results are shown at $z=3$ and $m=25$. 
Any change in astrophysical parameters will not alter the 
amplitude and shape of dark matter correlation function. 
However in principle they can change the luminosity 
dependent galaxy bias by simply changing the luminosity of a galaxy. 
Hence on large scales the astrophysical parameters will affect clustering 
predictions only through galaxy bias. 

As we can see explicitly in Fig.~\ref{fig:acf_z4_ratio_astro_2h}, 
the changes in astrophysical 
parameters produce negligible effects on the clustering on large scales. 
This is expected for $M_{agn}$ because of the following reason.
For $z=3$ we have chosen the parameter $M_{agn} \sim 0.8\times 10^{12}M_\odot$ 
so that the predicted LF matches with the observed data. In the plot we 
varied $M_{agn}$ between $0.5 \leq M_{agn}/10^{12} M_\odot \leq 1.1 $.
From Table~\ref{tab1} it is clear that, $M_{min}$, 
the minimum mass cutoff in the integral to calculate large scale bias 
(see Eq. ~\ref{eqn:biasL2}) is always less than $10^{12} M_\odot$.
Since the number density of halos decays exponentially with $\sigma^2(M)$ 
at these mass scales, the major contribution to $b_g(z,L_{th})$ 
in Eq. ~\ref{eqn:biasL2} comes from mass scales lower than $M_{agn}$. 
Therefore, we can conclude that the value of $M_{agn}$ 
and hence the AGN feedback 
used in our models will not be that sensitive to clustering on large scales. 

From Fig~\ref{fig:acf_z4_ratio_astro_2h} it is also clear that the 
change in $\kappa$ is not affecting the predicted clustering at large scales. 
We know that a change in $\kappa$ is not affecting the number density 
of halos of a particular mass. 
However a smaller $\kappa$ means the baryons are converted to stars
over a shorter timescale. Hence for a fixed $f_\ast$ 
(or the baryon fraction being converted to stars),
a smaller $\kappa$ leads to a increase in the SFR and 
increased luminosity.
This will shift the predicted total LF more or less along the luminosity axis. 
While fitting the observed luminosity function this shift with respect to 
our fiducial model is nullified by changing $f_\ast/\eta$. 
This means that even when the $\kappa$ is decreased and the 
star formation rate is enhanced, the 
observed luminosity of a galaxy hosted by a halo of a given mass is almost
unchanged, and hence the luminosity dependent 
large scale clustering is not significantly affected.

\subsubsection{Sensitivity to Cosmology}

\begin{figure}
\begin{center}
\includegraphics[trim=0cm 0cm 0cm 0cm, clip=true, width =8.0cm, height=9.0cm, angle=0]
{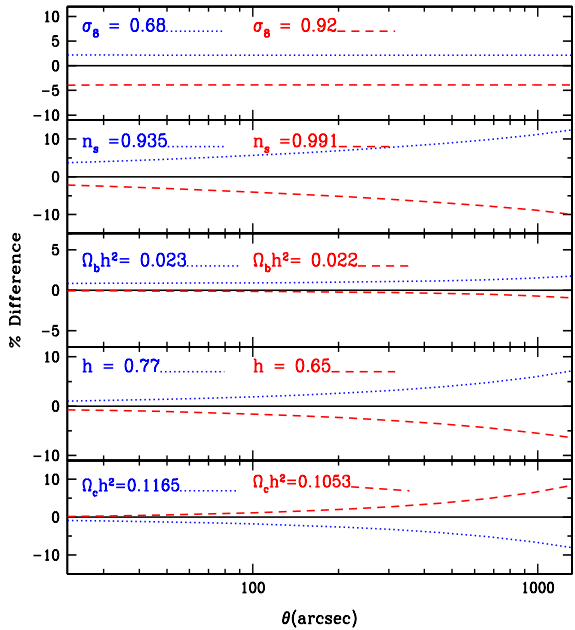} 
\caption{
The percentage change is angular correlation function when one varies the 
cosmological parameters around the fiducial model at $z=3$. 
Here the plots are for threshold apparent magnitude 25. 
In each case, $f_\ast/\eta$ values are varied to 
consistently reproduce the observed luminosity function. 
}
\label{fig:acf_z4_ratio_cosmo_2h}
\end{center}
\end{figure}

On large scales values of cosmological parameters 
can affect two point correlation function by 
changing the amplitude and shape of dark matter power spectrum  and also by 
modifying the halo mass function and halo bias and thereby altering 
the galaxy bias. To see these effects more clearly, we have plotted in 
Fig.~\ref{fig:acf_z4_ratio_cosmo_2h} the percentage differences in the 
clustering predictions with changes in cosmological parameters from their 
fiducial value. In this figure we varied the cosmological parameters 
within their $2\sigma$ limits determined by WMAP7 year data.
The parameter $f_\ast/\eta$ in each case is fixed by fitting  
the observed luminosity function. 
All the curves in Fig.~\ref{fig:acf_z4_ratio_cosmo_2h} are for $z=3$ and 
threshold apparent magnitude 25. 
The parameter values which are different from the fiducial 
value are shown in each panel of Fig~\ref{fig:acf_z4_ratio_cosmo_2h}. 
The percentage difference of clustering ranges from minimum of few $\%$ 
(for $\sigma_8$ and $\Omega_b$) and a maximum of $\sim$ 12\% (for $n_s$) 
for the assumed values of parameters, 
with notable differences being produced 
only on very large scales.

%In particular we see that the correlation functions on large scales 
%are not very sensitive to $\sigma_8$ and $\Omega_b$. 
%This is expected for $\Omega_b$ as any small change in 
%baryon density do not considerably alter the dark matter correlations, 
%halo mass function and halo bias. 
%On the other hand any change in $\sigma_8$ will significantly 
%modify all the above quantities.
%However 
We are particularly surprised why 
our models with various values of 
$\sigma_8$ predict almost same large scales clustering. 
In order to understand 
this
we consider two models A and B 
with different values of $\sigma_8$ given by $\sigma^A_8=0.68$ and $\sigma^B_8=0.80$. 
In Appendix A we have obtained an expression for the approximate ratio between 
galaxy correlation functions of models A and B on large scales (Eq. \ref{eqn:A6}). 
In particular we consider the clustering of galaxies with 
threshold apparent magnitude 25. 
From our analysis in Section~\ref{sec:xi_total} we can obtain the average mass of 
a galaxy, $M_{av}$, that can shine brighter than this magnitude threshold.  
For models A and B, the corresponding average masses are 
$M^A_{av}\sim 5.6\times10^{11} M_\odot$ and 
$M^B_{av} \sim 10^{12} M_\odot$ respectively. 
Using these in Eq. \ref{eqn:A6} we get the ratio between correlation 
functions on large scales for models A and B to be $1.05$. 
This is consistent with our model predictions shown in 
Fig.~\ref{fig:acf_z4_ratio_cosmo_2h}. 
Thus in our model various effects due to the change in $\sigma_8$ 
cancel each other to get similar large scale 
clustering of high redshifts LBGs. 

%The changes in all other cosmological parameters produce 
%notable differences in 
%the large scale clustering especially on very large scales. 
%These modifications to clustering on large scales 
%due to cosmology are quite interesting and promising. 
%We have already seen that on large scales changes in astrophysical 
%parameters do not strongly affect our model predictions of clustering as 
%long as $f_\ast/\eta$ is constrained by the observed LF.
%Moreover the percentage difference in correlation functions shown by 
%various models in Fig.~\ref{fig:acf_z4_ratio_cosmo_2h} appears to 
%increase with the separation scale. Although the current 
%observations cannot distinguish between these models, future surveys with large number 
%of sources detected in a large survey volume may distinguish between large scale  
%clustering predictions of high redshift LBGs with different cosmological parameters. 
%Hence we conclude that our model for large scale correlation function of 
%high redshift LBGs can be potentially used to constrain cosmological 
%parameters. 

%%%%%%%%%%%%%%%%%%%%%%%%%%%%%%%%%%%%%%%%%%%%%%%%%%%%%%%%%%%%%%%%%%%%%%%%%%%%%%%%%%%%%%%%%%%

\section{The correlation functions including the satellites}
\label{sec:xi_total}
In the previous sections we assumed that a halo can host at most one detectable galaxy. 
This assumption is not adequate to explain clustering at small angular scales, especially 
on scales smaller than virial radius. Each halo can in principle host multiple galaxies. 
A complete description of the distribution of galaxies inside a halo is called halo occupation 
distribution (HOD). Following the approach of ~\citet{kravtsov_04} 
we separate the central and satellite contributions to HOD 
(see also \cite{zheng_05,cooray_ouchi_06,conroy_wechsler_06}). 
That is the mean number of galaxies inside a dark matter halo can be written 
as $N_g(M) = f_{cen}(M)+N_{s}(M)$ where $f_{cen}(M)$ and $N_s(M)$ are respectively 
the mean number of central and satellite galaxies. 
We also assume the central galaxy to be situated at the center 
of the halo and satellite galaxies around it \citep{kravtsov_04}. 
In this approach the 1-halo term has contributions 
from the correlation between central-satellite and satellite-satellite pairs. 

In the framework of this approach, the the total correlation function can be 
written as \citep{cooray_sheth_02} 
\be
\xi_g(r) = \xi^{1h}_g(r) + \xi^{2h}_g(r)
\label{eqn:xi_total}
\ee
where each term on RHS has contributions from central as well as satellite galaxies.

\subsection{The HOD: a physical model}

Often the HOD is modelled in a parametrized form, motivated by simulations
or observations, and the value of the parameters are derived by fitting the
galaxy correlation function. Instead we adopt a more physical approach. We ask, 
given our model for galaxy formation, what is the expected mean occupation number of 
central and satellite galaxies of a given luminosity, in a given halo. These can 
then be used to compute the correlation function.

\subsubsection{The central galaxy occupation}
We begin by formally expressing the mean occupation number of central galaxies with a luminosity 
threshold $L_{th}$ inside a halo of mass $M$ at any redshift $z$. 
This is given by,
\bea
f_{cen}(L_{th},M,z)  &=& \f{\int\limits^\infty_{z} \df{dn(M,z_c)}{dz_c} 
                                     \Theta \left(L(M,z,z_c)-L_{th}\right) dz_c} 
                                  {\int\limits^\infty_{z} dz_c \df{dn(M,z_c)}{dz_c}}   \nonumber  \\
               &=& \f{\int\limits_{z_1}^{z_2} dz_c \df{dn(M,z_c)}{dz_c} }
                     {n(M,z)}.
           \label{eqn:fcen}
\eea
Thus, in our model $f_{cen}(L_{th},M,z)$ is the probability that a halo of mass $M$ at $z$ 
hosts a galaxy with brightness greater than $L_{th}$. 
Again 
the $\Theta \left(L_{th}(M,t(z)-t(z_c)\right)$ function ensures that only
those galaxies collapsing at $z_c$ and having $L >L_{th}$ at $z$ are
counted in the integral over $z_c$. As described earlier, $z_1(M,L_{th},z)$ and $z_2(M,L_{th},z)$ 
are the two redshifts at which a galaxy of mass $M$ has to be formed, 
so that it shines with an observed luminosity $L_{th}$ at $z$.  

In Fig. \ref{fig:Ng} we have plotted as thin lines, the average occupation 
number of central galaxies ($f_{cen}(M)$) as function of the mass of the 
parent halo, calculated using the above prescription. 
The results are shown at $z=3,4$ and $5$  for three different 
magnitude thresholds which are labelled in each panel. 
These curves are obtained using the fiducial 
set of model parameters (given in Section 2) that reproduce 
the observed luminosity function\footnote{Inclusion of the contribution 
of satellite galaxies slightly modifies the luminosity function and 
hence the best fit $f_*/\eta$ (see Section 4.4 ). 
In all the subsequent calculations we use
this new best fit $f_*/\eta$.}.
We can see that for a given redshift and  
limiting magnitude $f_{cen}(M)$ increases with the mass of the hosting halo 
to a limiting value of 1. Also note that the mean occupation number 
of central galaxies drops to zero below some mass scale $M_{min}$. 
This value of $M_{min}$ is almost the same as that given in Table.~\ref{tab1};
the change is due to the minor corrections to $f_*/\eta$ we apply 
to take care of the satellite contribution to the LF
in Section 4.4.
In Table~\ref{tab2} we give the values of $M_{min}$ for 
all the three redshifts and limiting magnitudes.
For example, at $z=4$ for apparent magnitudes 25, 25.5 and 26,  
$M_{min}= 5.7\times 10^{11}$, $3.4\times10^{11}$ 
and $2.1\times 10^{11}~M_\odot$ respectively. 
Many of the semi analytical models for clustering of low redshift galaxies 
\citep{zehavi_zheng_11,more_09} use a step like function with smooth 
cut-off profile for $f_{cen}(M)$. One can see from  Fig. \ref{fig:Ng} 
that such a profile 
for $f_{cen}(M)$ naturally arises from our simple physical model of star formation. 
Infact the smoothness of the cut-off seen in the function $f_{cen}(M)$ at the low mass 
end for our models, is due to the scatter in the M-L relationship.   

It is of interest to calculate the average value of $f_{cen}$ for halos
above the threshold mass $M_{min}$, defined as 
\be
\langle f_{cen} \rangle(L_{th},z) = 
\frac{\int_{M_{min}}^\infty dM \ f_{cen}(M,z) n(M,z)}
{\int_{M_{min}}^\infty dM n(M,z)} .
\label{fcenav}
\ee
We give the value of $\langle f_{cen} \rangle$ in Table~\ref{tab2}
for different magnitude thresholds and redshifts. 
One can see from Table~\ref{tab2} that $\langle f_{cen} \rangle$ is generally
of order $0.4$; or $40\%$ of halos above $M_{min}$ can typically
host luminous LBGs, for any given luminosity threshold and redshift.  
This number, which in our model only depends on $f_\ast/\eta$
obtained by a fit to the LFs, is comparable to the
duty cycle values preferred by \citet{lee_09,lee_ferguson_12}.
 
Finally we also give in Table~\ref{tab2}, the average mass $M_{av}$ 
of a halo hosting the central LBG, 
for each redshift and luminosity threshold. This
average mass is defined by, 
\be
M_{av} = \frac{\int_0^\infty dM \ M f_{cen}(M,z) n(M,z)}
{\int_0^\infty dM f_{cen}(M,z) n(M,z)} .
\label{mav}
\ee
At $z=4$ and for apparent magnitudes 25, 25.5 and 26,  
$M_{av}= 10^{12}$, 
$6.2\times10^{11} $ 
and $3.9\times 10^{11} ~M_\odot$ respectively.
At $z=5$ these masses are smaller by a factor $\sim1.4$ while
at $z=3$, $M_{av}$ is larger by a similar factor.

We also note that $b_g(k,L_{th},z)$ in Eq.~(\ref{eqn:biasL2}) 
can be rewritten
in terms of $f_{cen}(L_{th},M,z)$ as,
\bea
b_g(k,L_{th},z)  =&& \f{1}{n_g(L_{th},z)}\int^\infty_0 dM~ n(M,z) b(M,z) \times  \nonumber \\
&& f_{cen}(L_{th},M,z) u(k,M,z)
\label{eqn:biasL3}
\eea 
which is the standard formula for the galaxy bias, excluding the satellite
contribution. 
 
\subsubsection{The satellite galaxy occupation}
In order to get an estimate of the number of subhalos and thereby the number of 
satellite galaxies hosted by a halo of mass $M$, we use the conditional or progenitor 
mass function. The conditional mass function 
gives the comoving number density of subhalos of mass $M_s$ which formed 
at $z_s$ inside a region containing a mass $M$ 
(or comoving volume $V$) that have a non-linear over density $\delta$ at $z_p$. 
It is given by \citep{mo_white_96, cooray_sheth_02}
\be
\bar n(M_s,z_s|M,\dl,z_p)dM_s  = \f{\rho_m}{M_s} \nu_{10} f(\nu_{10}) \frac{d\nu_{10}}{\nu_{10}}
\label{eqn:cmassfn}
\ee
where 
\be
\nu_{10} = \f{\left(\dl_{c}(z_s) - \dl_l(\dl,z_p)\right)^2}{\sigma^2(M_s)-\sigma^2(M)}.
\ee
Here, $\dl_l(\dl,z_p)$ is the linear density contrast at redshift 
$z_p$ corresponding to non-linear over density $\dl$ and $\rho_m = \Omega_m \rho_{c}$ is 
the back ground density of baryons and cold dark matter. 
It should be noted that the formation epoch of the subhalo 
should precede the formation epoch of the parent halo ($z_s>z_p$) 
and mass of the subhalo should be always smaller than that of the 
parent halo ($M>M_s$). 
The function on the RHS of Eq.~(\ref{eqn:cmassfn}) has the same form as the 
unconditional mass function but with $\dl^2_{c}(z_s)/\sigma^2(M)$ being replaced by 
$(\dl_{c}(z_s) - \dl_l(\dl,z_p))^2/(\sigma^2(M_s)-\sigma^2(M))$. 
We use Sheth-Tormen form for the RHS in Eq.~(\ref{eqn:cmassfn}),
in our calculations (see Section 3.3 of ~\citet{cooray_sheth_02} for more details).
The total number of halos of mass $M_s$ which formed 
at $z_s$ inside the volume containing mass $M$ is
\be
N(M_s,z_s|M,\dl,z_p)dMs = (M/\rho_m) \bar n(M_s,z_s|M,\dl,z_p)dMs,
\ee
where we have multiplied $\bar n$ by the comoving volume $(M/\rho_m)$.
   
\begin{figure}
\includegraphics[trim=0cm 0cm 9cm 0cm, clip=true, width =8.0cm, height=12cm, angle=0]
{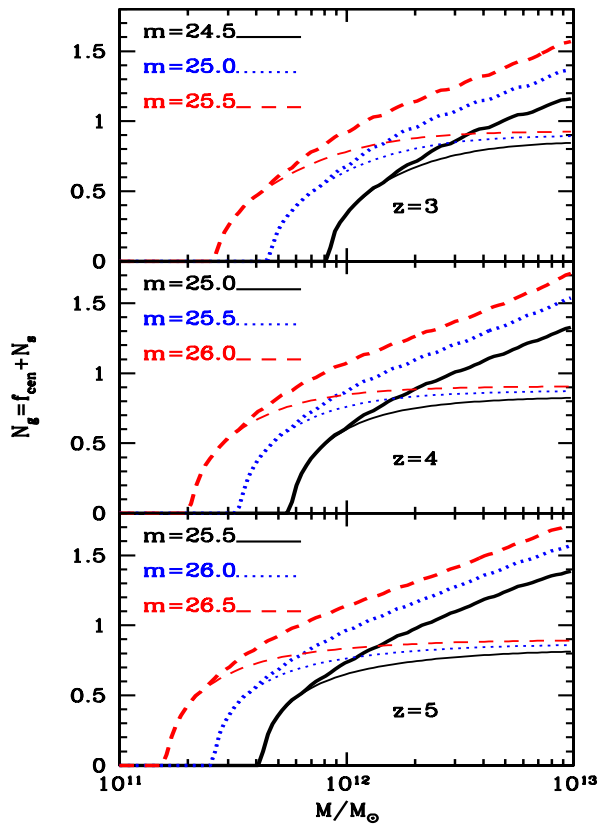} 
\caption{
The  halo occupation distribution, $\langle N_g(M) \rangle$, 
as a function of the mass of the 
hosting halo, as predicted by our models at various redshifts for three limiting magnitudes. 
In each panel the thin curves corresponds to $f_{cen}(M)$ where as the thick curves give 
the total occupation $N_g(M) = f_{cen}(M)+N_s(M)$. The redshifts and apparent magnitudes 
are labelled in each panel. The curves are obtained using a fiducial 
set of model parameters (given in Section 2) that reproduce the observed 
luminosity function. In addition to that, to obtain the satellite occupation ($N_s(M)$),  
we have adopted $\Delta t_0 = 1.6 t_{dyn}, 1.5 t_{dyn}$ and $1.4 t_{dyn}$ 
at $z=3, 4$ and $5$ respectively (see text for details). 
}
\label{fig:Ng}
\end{figure}

\par

Our aim is to find the number of satellite galaxies of a particular luminosity 
inside a dark matter halo of mass $M$. To calculate this we assume that the 
region containing mass $M$ 
%(Eq.~(\ref{eqn:cmassfn})) 
is a collapsed dark 
matter halo at $z$. Thus in Eq.~(\ref{eqn:cmassfn}) we take 
$\dl_l(\dl,z_p) = \dl_c(z_p)$ \citep{cooray_sheth_02}. 
In this limit the conditional mass function gives the 
number density of satellite halos in the mass range 
$M_s$ and $M_s+dM_s$ at $z_c$ inside a halo of mass $M$ that collapsed at $z$. 
We choose the time derivative of $N(M_s,z_c|M,\dl,z)$ as the formation rate 
of subhalos of mass $M_s$ inside a big halo of mass $M$. 
%This can be justified by the the assumption 
We also assume
that even though dark matter 
halos are formed and destroyed inside the over dense cell of mass $M$, 
the satellite galaxies formed in these subhalos have survived 
and can be observed. 

We further assume that no subhalo should be formed very close to the 
formation epoch of parent halo. More precisely if 
$t_p=t(z_p)$ is the age of the 
universe when a parent halo collapsed then all the subhalos 
formed inside that parent halo within a time interval $\Delta t_0$ prior  
to $t_p$ do not host a satellite galaxy; rather 
they will be part of the parent halo itself. 
Thus we assume $t_p-t_s \geq \Delta t_0$ where $t_s=t(z_s)$ is the age of 
the universe when the sub halo formed inside a parent halo.
In our models we vary $\Delta t_0$ 
as a free parameter. Also we expect that it is of the order of dynamical 
time scale $t_{dyn}$ of the parent halo. It should be noted that 
the time scale $\Delta t_0$ corresponds to a redshift interval 
$\Delta z_0$ which is a function of the redshift of 
collapse of the parent halo\footnote{We have also explored 
an alternate scheme whereby halos of satellite galaxies 
must have a mass smaller than the parent halo by a 
factor $f_m$ (same as the parameter $\varphi$ in 
\cite{lee_09}). Since smaller mass halos typically collapse earlier
than larger masses, the parameter $\Delta t_0$ used here
achieves a similar goal to $f_m$; although it does not exclude similar
but slightly smaller mass satellites in the main parent halo.}.

In principle, a parent halo seen at $z$ could itself have collapsed
at a slightly earlier redshift $z_p > z$. However, it turns out 
(see Fig.~\ref{fig:Ng}), 
that only massive halos with $M \geq 10^{12} M\odot$ have significant
probability of hosting an observable satellite galaxy.
Such massive halos are more likely to collapse close to the redshift
of observation z itself (as their formation rate will rapidly fall with
increasing $z_p$). Thus for simplifying our computations of the satellite
occupation, we will assume
that the collapse redshift of the parent halo is also the redshift
of observation; i.e $z_p = z$. Of course for computing $f_{cen}$ we 
do not make any such approximation.
%We will show that the typical mass of the halos that can host 
%satellite galaxies satisfying the luminosity threshold condition is 
%$M \geq 10^{12} M\odot$. 
%At high redshifts $(z>3)$ most of these parent 
%halos have collapsed close to the the redshift of observation. 
%So in this Section for computational purpose we approximate the redshift of 
%collapse 
%of the parent halos halos ($z_p$), 
%that can host visible satellite galaxies, to be the redshift 
%of observation ($z$) itself (ie $z_p = z$).

In the following discussions we also assume that star formation 
models for the central and satellite galaxies are identical 
(given by Eq.~(\ref{eqn:SFR})). This is similar to the assumption used by 
previous authors \citep{lee_09,berrier_06,conroy_wechsler_06} where feedback 
due to halo merging processes are ignored. 

We first compute the number of satellite galaxies of 
magnitude in the range $M_{AB}$ and $M_{AB}+dM_{AB}$ 
inside a halo of mass $M$. This is given by,
\bea
\varphi(M,M_{AB}, z ) dM_{AB} = \int\limits_{z-\Delta z_0}^\infty 
                           \f{dN(M_s,z_s|M,z)}{dz_c}  \times \nonumber \\
                  \frac{dMs}{dL_{1500}} ~\frac{dL_{1500}}{dM_{AB}}~ dM_{AB} ~dz_c. 
		   \nonumber \\ 
\eea
It should be noted that $\varphi(M,M_{AB},z) dM_{AB}$ is not the luminosity 
function of satellite galaxies, but the total number of satellite galaxies 
in a magnitude range $M_{AB}$ and $M_{AB} + dM_{AB}$ inside a halo of mass $M$. 
The average number of satellites, $N_s$, with luminosity greater than $L_{th}$ in 
halo of mass $M$ at redshift $z$ can be computed as 
\be
\langle N_s|L_{th},M,z)\rangle =  \int^\infty_{L_{th}} \varphi(M,L,z) dL
\ee
where the luminosity $L$ is related to $M_{AB}$ in the standard manner.

\subsubsection{The total halo occupation}
We have plotted in Fig. \ref{fig:Ng} the total occupation number of  
galaxies, $N_g(M)=f_{cen}(M)+N_s(M)$, as a function of the 
mass of the parent halo. These curves are shown in thick lines for  
at $z=3,~4$ and $5$ and for three luminosity thresholds. 
Thin lines correspond to occupation number of central galaxies. 
In order to obtain $N_s(M)$, apart from the fiducial model parameters 
that fit the observed luminosity function, 
we have adopted $\Delta t_0 = 1.6 t_{dyn}, 1.5 t_{dyn}$ and $1.4 t_{dyn}$ 
at $z=3, 4$ and $5$ respectively. These values of $\Delta t_0$ are chosen  
as they are later used in Section 5.1 for
explaining the small angular scale clustering.
We find that for each limiting magnitude the mean number of satellites 
is a monotonically increasing function of hosting halo mass. 
As an example at $z=4$ for a halo of mass $2\times10^{12} M_\odot$ 
the mean satellite occupation numbers are respectively 0.14, 0.27 and 0.38 
at threshold apparent magnitudes of 25, 25.5 and 26. 
For a bigger halo of mass $5 \times 10^{12} M_\odot$ these numbers change to 
0.34, 0.48 and 0.61 for the same limiting magnitudes at the same redshift.

We have also computed the average value of $N_s$ in a manner
similar to calculating $\langle f_{cen}\rangle$ in Eq.~(\ref{fcenav}).
The results are given in the last column of Table~\ref{tab2}.
We see from the table that typically $\langle N_s\rangle$ is only about
5-10\% of $\langle f_{cen}\rangle$.
Thus the average number of detectable satellites, in a halo, 
is typically much less
than unity. This implies that all halos do not necessarily host an
additional detectable satellite galaxy; however some small 
fraction of them do
and it is these pairs of LBGs which contribute to
the small scale clustering. Such a conclusion has
also been arrived at by \cite{conroy_wechsler_06}
using numerical simulations.

It is also of interest to ask for a characteristic mass $M_P$
of a parent halo which is hosting a detectable satelite galaxy.
We define this as follows:
\be
M_{P} = \frac{\int_0^\infty dM \ M N_s(L_{th},M,z) n(M,z)}
{\int_0^\infty dM N_s(L_{th},M,z) n(M,z)}
\label{mparent}
\ee
This is also given in Table~\ref{tab2}. 
For apparent magnitudes 25, 25.5 and 26, and $z=4$, 
$M_P=1.9\times 10^{12} M_\odot$, $1.2\times10^{12}M_\odot$ and 
and $7.4\times 10^{11}M_\odot$ respectively.
For $z=3$ the corresponding $M_P$ is $\sim 1.5$ times larger,
while for $z=5$, $M_P$ is smaller by a factor $\sim1.4$.
This also means that the characteristic mass of 
parent halos hosting detectable satellites
is $\sim 2$ times larger than the mass
of the halo hosting a central LBG. 

Note that the mean satellite number obtained in our physically 
motivated models can be fit asymptotically by a power law of the form 
$N_s(M) \propto M^{\alpha}$, with $\alpha \sim 0.6-0.9$.
Interestingly this is similar to the parametrized form 
$N_s(M)$ used in \cite{hamana_06}, who also adopt a similar slope.
Moreover, the form of HOD that we derive from our physically motivated
model (and shown in Fig.~\ref{fig:Ng}) is also similar to 
that obtained by \cite{conroy_wechsler_06} from their models
of HOD using N-body simulations combined with
a prescription of associating mass to light 
(see Figures 8 and 10 of their paper). 

\subsection{The one halo term}
We can now compute the 1-halo term using the standard assumption 
that radial distribution of satellite galaxies inside 
a parent halo follow the dark matter 
density distribution \citep{cooray_sheth_02}. 
For the present calculations we use the NFW form 
for the dark matter density distribution. In this case the 1-halo term is given by 
(\citet{tinker_weinberg_05,cooray_ouchi_06}; 
see also \citet{skibba_sheth_09})
\bea
\xi^{1h}(L_{th},R,z) = \f{1}{(n_g^{T})^2}\int dM~ n(M,z) \times  \Big[ 2  f_{cen}(L_{th},M,z)  \nonumber \\ 
   \langle N_s|L_{th},M,z\rangle   \f{\rho_{NFW}(R|M)}{M} + \nonumber \\
    \langle N_s(N_s-1)|L_{th},M,z  \rangle \f{\lambda_{NFW}(R|M)}{M^2} \Big].
\label{eqn:xir-1h}
\eea
Here, 
\be
n^T_g(L_{th},z)= \int dM n(M,z)(f_{cen}(M,z)
+\langle N_s|L_{th},M,z\rangle)
\ee 
is the total number density of galaxies which includes both the central and
satellite galaxies. 
Further $\rho_{NFW}$ is the NFW profile of dark matter density inside a 
collapsed halo and $\lambda_{NFW}(r|M)$ is the convolution of this density 
profile with itself \citep{sheth_hui_01}. 
The NFW density profile is given by 
\be
\rho_{NFW}(M,R) = \df{4\rho_s}{(R/R_s)(1+R/R_s)^2}
\ee
where the $\rho_s$ and $R_s$ are characteristic density and radius respectively.
%$\rho_s$ of a halo of mass $M$ 
%is related to the characteristic radius by $M = 4\pi R^3_s \rho_s/3$. 
The ratio of the virial radius and the characteristic radius of the halo 
is defined 
as the concentration parameter ($c=R_{vir}/R_s$).
For the halo concentration parameter we use the fitting functions given 
by \cite{prada_klypin_12} (Eq.~(14-23) of their paper). 
These fitting functions of concentration parameter 
at high redshifts, unlike earlier fits, flatten and then upturn with increasing 
mass. We also check the sensitivity of our results to other expressions for 
concentration parameter. 
Also following the N-body simulations and semi-analytical 
models \citep[eg.~][]{kravtsov_04,zheng_05}, we assume that the number 
of satellites inside a parent halo forms a Poisson distribution. 
Thus we have $\langle N_s(N_s-1)\rangle =  N^2_s$.

\subsection{The two halo term}
The correlation between satellite-satellite and satellite-central 
pairs located in different halos also modifies the 2-halo term. 
In this case the expression for scale dependent galaxy bias in 
Eq.~(\ref{eqn:biasL3}) can be modified by adding in the contribution from
satellite galaxies. Note that each halo of mass $M$ has $N_s(L_{th},M,z)$ 
satellite galaxies, brighter than $L_{th}$. Thus the number density 
of satellite galaxies associated with a halo in any mass interval 
$M$ and $M+dM$ is $n(M,z)N_s(L_{th},M,z) dM$. 
These satellites are residing in a parent 
halo with a large scale bias $b(M,z)$. Therefore, the average galaxy 
bias obtained by adding the contributions of both the central and satellite 
galaxies is,
\bea
b_g(k,z,L_{th}) = \f{1}{n^T_g(L_{th},z)} \int dM n(M,z) b(M,z) \nonumber \\
             ~~~~ \Big[f_{cen}(L_{th},M,z) +\langle N_s|L_{th},M,z)\rangle \Big] u(k,M,z). 
\label{eqn:bias_total}
\eea
%Here $n^T_g(L_{th},z)= \int dM n(M,z)(f_{cen}(M,z)+\langle N_s|L_{th},M,z\rangle)$ 
%is the total number density of galaxies which includes both the central and
%satellite galaxies. 
If halos do not host any satellite galaxies, 
ie $ \langle N_s|L_{th},M,z)\rangle = 0$, this expression reduces back 
to Eq.~(\ref{eqn:biasL3}). We can now compute the two halo term by 
substituting Eq.~(\ref{eqn:bias_total}) into Eq.~(\ref{eqn:pk}) and 
using Eq.~(\ref{eqn:xir-2h}).

\begin{table}
\tabcolsep 4.6pt
\caption{
Physical parameters constrained using observed luminosity function of 
high-$z$ LBGs after incorporating the satellite galaxies in our semi-analytical models. 
These parameters are $f_\ast/\eta$, an indicator of the light to mass ratio at 
any redshift, $M_{min}(L_{th},z)$, the minimum mass of a galaxy that can shine brighter 
than a given luminosity threshold $L_{th}$ at redshift of observation, 
$M_{av}(L_{th},z)$ is the average mass of a dark matter halo that can host a central LBG   
satisfying the luminosity threshold $L_{th}$,  $M_{P}(L_{th},z)$ the characteristic 
mass of a parent dark matter halo that hosts satellite LBGs which satisfy 
the luminosity threshold $L_{th}$, $\langle f_{cen}\rangle$, the average central 
occupation and  $\langle N_{s}\rangle$, the average satellite occupation. 
}
%\label{table1}
\begin{tabular}{cccccccc}\hline
$z$ &$f_\ast/\eta$ &$m$ \T \T  \B\B \B\B &$\f{M_{min}}{10^{11} M_\odot}$ &$\f{M_{av}}{10^{11} M_\odot}$ 
 &$\f{M_{P}}{10^{11} M_\odot}$ &$\f{\langle f_{cen} \rangle}{10^{-2}}$ &$\f{\langle N_s \rangle}{10^{-2}}$\\
\hline \TT  
\T &        &24.5  &$8.2$  &$15.7$ &$30.0$  &40.1 &2.2\\
 3  &0.040  &25.0  &$4.6$  &$9.2$  &$18.0$  &39.1 &3.0\\
\B &        &25.5  &$2.7$  &$5.6$  &$11.1$  &39.2 &3.8\\

\T &        &25.0  &$5.7$  &$10.0$  &$18.8$ &40.0 &2.5\\
 4  &0.035  &25.5  &$3.4$  &$6.2$  &$11.7$  &40.3 &3.0\\
\B &        &26.0  &$2.1$  &$3.9$  &$7.4$   &42.6 &3.8\\

\T &        &25.5  &$4.2$  &$7.0$  &$12.2$  &38.7 &2.6\\
 5  &0.030  &26.0  &$2.6$  &$4.5$  &$7.7$   &41.5 &3.3\\
\B &        &26.5  &$1.6$  &$2.9$  &$5.0$   &42.5 &3.8\\

\hline
\end{tabular}
\label{tab2}
\end{table}

\subsection{The total luminosity function}

\begin{figure}
\begin{center}
\includegraphics[trim=0cm 0cm 0cm 0cm, clip=true, width =8.0cm, height=7.5cm, angle=0]
{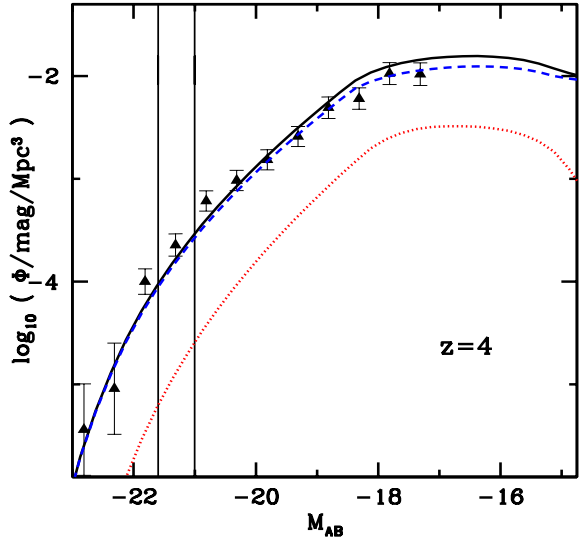} 
\caption{
The best fit luminosity function of LBGs at $z=4$ (solid-black) 
including the contribution from satellite galaxies. Also shown 
is the individual luminosity functions due to central (dashed-blue) 
as well as satellite (dotted-red) galaxies. The vertical (solid-black) 
lines are drawn at absolute magnitudes -21.0 and -21.6.
}
\label{fig:LF_both}
\end{center}
\end{figure}

One may wonder
if the additional satellite galaxies inside a halo can modify the luminosity 
function. In this section we compute the contribution to galaxy luminosity 
function from visible satellite galaxies. The total luminosity function of 
LBGs at any redshift can be written as, 
\be
\Phi(M_{AB}, z ) = \Phi_{cen}(M_{AB}, z ) + \Phi_{sat}(M_{AB}, z ).
\ee
In section \ref{section:SFR} we computed the luminosity function due to 
central galaxies. The contribution to luminosity function due to 
satellite galaxies can be written as,
\be
\Phi_{sat}(M_{AB}, z ) dM_{AB} =  \int dM \f{\rho_m}{M} \varphi(M, M_{AB},z) n(M,z).
\ee

In Fig. \ref{fig:LF_both} we have plotted the luminosity functions 
of both central and satellite galaxies separately at $z=4$. 
It is clear that the number density of satellite galaxies 
(red dotted curve) at any given $M_{AB}$ is roughly 
an order of magnitude less compared to that of 
central galaxies (blue dashed curve) at any luminosity bin. 
Thus the contribution to total luminosity function 
(solid black curve) due to satellite galaxies is negligible 
at any given luminosity.
However, these satellites can still contribute to the
small scale clustering. As discussed earlier, the typical mass $M_P$ 
of parent halos hosting satellite galaxies is roughly 2 times
the average halo mass $M_{av}$ of a detectable LBG itself. 
As the luminosity roughly scales with halo mass, the 
central galaxy in the parent halo could then be about 2
times (or about $0.60$ magnitudes) brighter. From 
Fig.~\ref{fig:LF_both}, we can see that
satellite galaxies with absolute magnitude say $-21$, 
in a parent halo of mass $M_P$, will be about 
a third as abundant as the central galaxies, with 
absolute magnitude $-21.6$, in them.
Such pairs which will then occur in every third parent
halo of mass $M_P$ (hosting a detectable galaxy) 
will contribute significantly to the small
angular scale clustering.

\section{Comparisons with observations}
\begin{figure*}
\includegraphics[trim=0cm 0cm 0cm 0cm, clip=true, width =17cm, height=15cm, angle=0]
{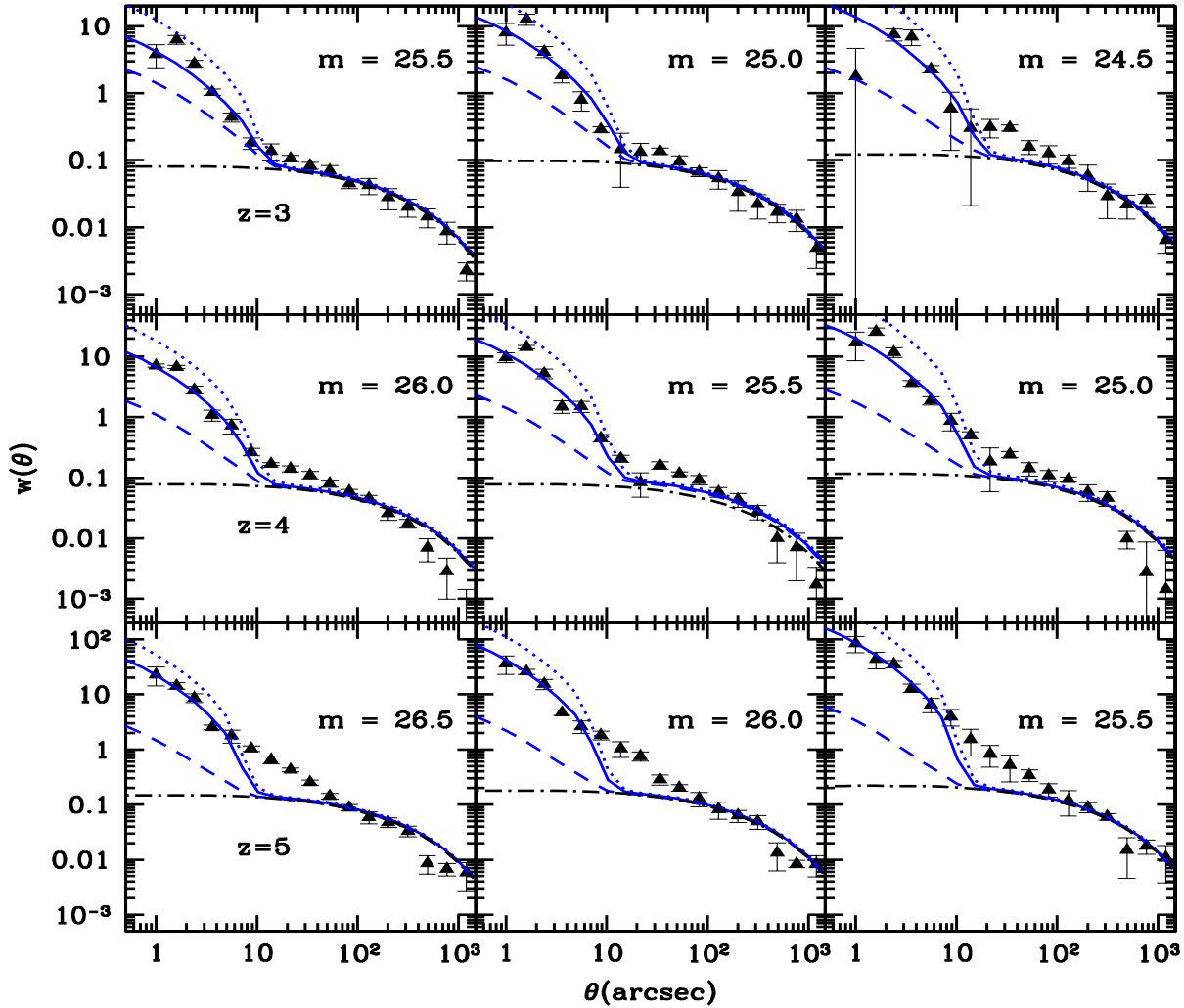} 
\caption{The angular correlation function of LBGs, taking into account 
the one halo and the two halo terms, at redshifts 3, 4 and 5 
for various limiting magnitudes. Each row corresponds to a particular redshift, 
which is labelled in the first panel of that row. 
In each row there are three panels showing the clustering predictions for 
galaxies with three threshold magnitudes, that are also labelled 
in the respective panels. 
Our fiducial model predictions of galaxy angular correlation 
functions with $\Delta t_0 = 1.6 t_{dyn}, 1.5 t_{dyn}$ and $1.4 t_{dyn}$ 
at $z=3, 4$ and $5$ respectively are shown in blue solid lines. 
The dotted and dashed curves are 
for $\Delta t_0 = t_{dyn}$ and $2 t_{dyn}$ respectively. 
The black dash-dotted lines show the correlation functions computed 
without satellite contribution (see Fig. ~\ref{fig:acfz_lin}). 
The data points and error bars shown by solid black 
triangles are from \protect \cite{hildebrandt_09_acf}.
% and those 
%shown by red squares are from 
%\protect \cite{ouchi_hamana_05_acf}.
}
\label{fig:acfz_full}
\end{figure*}
\subsection{The total correlation functions}
The total two point correlation function of galaxies at any scale is the sum 
of the contributions from one halo and two halo 
terms (as given in Eq.~(\ref{eqn:xi_total})). 
This can be converted to the angular correlation functions 
%can be estimated, 
using Eq.~(\ref{eqn:limber}).
% where one has to use 
%this total galaxy two point correlation functions, 
%as described in previous sub-sections. 
We again adopted the same fiducial model 
with $\kappa=1.0$ and three different $M_{agn}$ 
values, $0.8\times 10^{12}M_\odot$, $1.5\times 10^{12}M_\odot$, 
and $3.0\times 10^{12}M_\odot$ at redshifts 3, 4 and 5 respectively
($\kappa$ and $M_{agn}$ are described in Table~\ref{tab0}). 
The cosmological parameters are kept to their fiducial value. 
We also use $\Delta t_0 = 1.6 t_{dyn}, 1.5 t_{dyn}$ and $1.4 t_{dyn}$ 
at $z=3, 4$ and $5$ respectively for all the models, 
but test the sensitivity of the results to this choice.
This free parameter will correspond to different redshift 
intervals $\Delta z_0$ at $z=3$, $4$ and $5$.  
Various parameters of our fiducial model are tabulated in Table. ~\ref{tab3}.

The total angular correlation functions computed using our prescription for 
three redshifts and three threshold magnitudes are overplotted on the 
observed data in Fig.~\ref{fig:acfz_full}. 
In this figure the blue solid curves are 
our predictions of total galaxy angular correlation functions
for the fiducial model.
The dotted and dashed curves are 
for $\Delta t_0 = 1.0~t_{dyn}$ and $2~t_{dyn}$ respectively. 
We see that adopting the fiducial $\Delta t_0$ values quoted above
provides a better fit to the data at small angular scales, for all 
redshifts and magnitudes. A smaller (larger) $\Delta t_0$ leads generically to
an excess (deficit) of small angular scale clustering.
The black dash-dotted lines show the contribution to correlation function 
due to central galaxies alone (computed in Section ~\ref{sec:xi_linear} 
and given in Fig. ~\ref{fig:acfz_lin}). 
The satellite galaxies contribute negligibly to clustering at
large angular scales, as suggested earlier in Section~3.
The remaining contribution, which dominates especially at
small angular scales ($\theta <10''$), is mainly due to the
one halo term. It is also clear that the one halo term 
does not affect clustering on large angular scales ($\theta >10''$).
Note that the one halo and two halo contributions are distinctly
seen in the observed data for $z=3$ and $z=4$ as predicted
by our models. But these distinct contributions are not
so clearly seen in the $z=5$ data.

\begin{figure*}
\includegraphics[trim=0cm 6cm 0cm 0cm, clip=true, width =17.0cm, height=10.5cm, angle=0]
{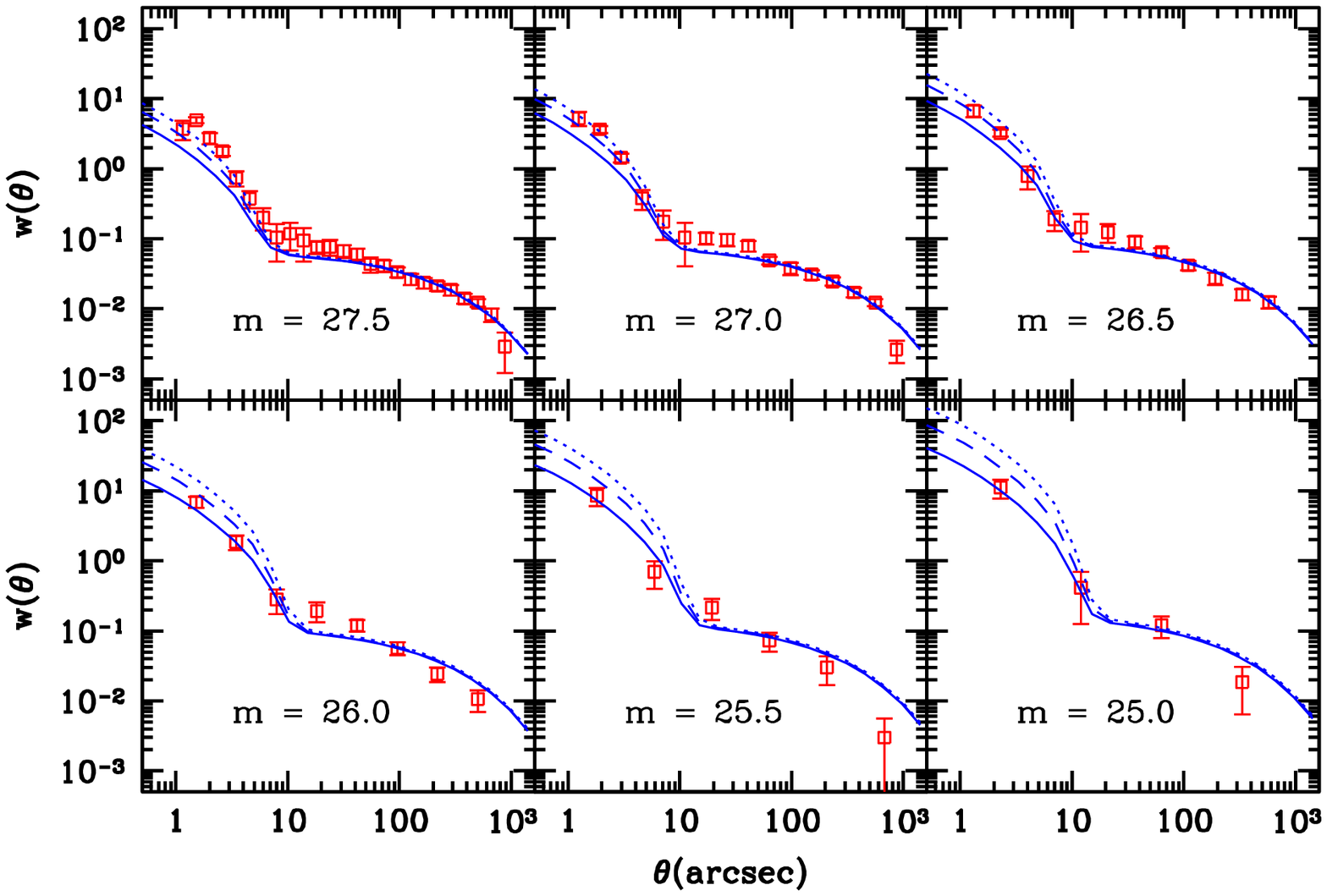} 
\caption{The angular correlation function of LBGs, taking into account 
the one halo and the two halo terms, at redshift 4 for various limiting magnitudes. 
Our fiducial model predictions of galaxy angular correlation 
functions with $\Delta t_0 = 1.5~t_{dyn}$ is 
shown in blue solid line. The dotted and dashed blue curves are obtained respectively by 
adopting $\Delta t_0$ to be $1.0~t_{dyn}$ and $1.25~t_{dyn}$.
The data points and error bars shown by red squares 
are from \protect \cite{ouchi_hamana_05_acf}.
}
\label{fig:acfz4_ouchi}
\end{figure*}

From the figure it is clear that our simple physical model gives a reasonable 
fit to the observed angular correlation functions 
at large ($\theta>80''$) and small ($\theta < 10''$) angular scales.
This is true at all magnitude thresholds and at all redshifts.
Moreover at $z=3$ the the predicted angular correlation
functions fits the observed data reasonably well at all angular 
scales (including the range $10''<\theta<80''$), 
for magnitude thresholds 25.0 and 25.5. 
However, for the most
luminous galaxies at $z=3$ (with $m < 24.5$), and 
for redshifts 4 and 5,
there is a discrepancy in the intermediate scales between 
the theoretically predicted correlation function and observed 
data points. 

In order to explore these issues further, we now consider the LBG clustering
measurements at $z=4$ by \citet{ouchi_hamana_05_acf}, which extends
to  much fainter galaxies with $m\le 27.5$.
We show in Fig.~\ref{fig:acfz4_ouchi} our model predictions 
of correlation functions at $z=4$ along with observed data (red 
squares) given by \cite{ouchi_hamana_05_acf}.  
In this figure the predicted angular correlation 
functions with $\Delta t_0 = 1.5~t_{dyn}$, that fits the
\citet{hildebrandt_09_acf} data, is 
shown in blue solid lines. The dotted and dashed blue curves are obtained 
respectively by adopting $\Delta t_0$ to be $1.0~t_{dyn}$ and $1.25~t_{dyn}$.
Note that here we have used the appropriate 
redshift distribution of galaxies 
($N(z)$ in Eq.\ref{eqn:limber}) as in \citet{ouchi_hamana_05_acf}. 

It is clear from Fig.~\ref{fig:acfz4_ouchi}, 
that for the fainter galaxy sample ($m \geq 26.5$), 
our physical model predicts a better fit to observed $w(\theta)$ 
even at the intermediate scales ($10'' < \theta < 80''$).
Note that this was also the case for the faintest LBGs at $z=3$
(see Fig.~\ref{fig:acfz_full}).
The typical mass of dark matter halo hosting a galaxy increases
with its luminosity.
For example, we find that 
the minimum mass of the dark matter halos that can host a galaxy 
with $m<27.5$ is $4.8\times 10^{10} M_\odot$ compared 
to $5.7\times 10^{11} M_\odot$ for a galaxy with $m<25.0$.
The higher order (quasi-linear) corrections to 
dark matter halos bias \citep{cooray_sheth_02} 
is expected to increase with the mass of the halos 
\citep{scannapieco_barkana_02,iliev_scannapieco_03}.   
This then suggests that the discrepancy at intermediate scales 
seen in Fig.~\ref{fig:acfz_full} could be due to 
missing quasi-linear (higher order) bias in our models.

One can also see from  Fig.~\ref{fig:acfz4_ouchi} that, 
for fainter galaxies, a smaller value of $\Delta t_0$ gives a 
better fit to the small scale clustering 
($\theta < 10''$). For example, for LBGs with $m= 25.0-26.0$,
the fiducial value $\Delta t_0 = 1.35 t_{dyn}$ gives a good fit 
to the observed small scale clustering. However, 
for fainter LBGs with $m=27.5$, one requires a smaller  
$\Delta t_0 \sim 1.0 t_{dyn}$ to obtain a good fit on
small angular scales. Since the typical masses of fainter galaxies 
are smaller than that of brighter galaxies, this suggests that the 
parameter $\Delta t_0$ could be a function of the masses of the 
parent and satellite galaxies. Such a mass dependence could be
an issue to probe further in the future when the small angular scale 
clustering data becomes more extensive. It could for example reflect the
fact that the SFR of satellites is dependent on the 
properties of the parent halos.

Note that our model predictions of the large angular scale clustering 
is almost free of parameters, once we fix the cosmology
and $f_\ast/\eta$ from fitting the LFs of LBGs.
We have given in Table~\ref{tab3}, the 
reduced chisquare 
$\chi_\nu^2$ obtained by comparing
the predicted angular correlation function in the range
$80'' < \theta < 600''$ with the observed data given by 
both \citet{hildebrandt_09_acf} ($\chi_\nu^2(H)$) 
and \citet{ouchi_hamana_05_acf} ($\chi_\nu^2(O)$).
One can also see from the values of $\chi_\nu^2$ given in the table,
that our model predictions reasonably fits all the data, 
except for the $m=26.5$ sample at $z=5$
(where there is one discrepant data point).

Note that the 
discrepancy that we have found between 
model predictions for the brightest LBGs
and the observed clustering data at intermediate scales, is also
present in some of the previous works 
(see Fig. 9 and 10 of \citet{lee_09}
and Fig. 5 and 6. of \citet{hamana_04}).
As we have argued above this is likely to be due to
quasi-linear corrections to the bias for the brightest LBGs.
In addition to these higher order corrections to the bias, 
the missing clustering power at intermediate scales
could also be due to 
(i) uncertainties related to distribution of satellite galaxies that is 
assumed to follow the NFW density profile of dark matter halos, 
(ii) the non-linear corrections to dark matter power spectrum 
\citep{mo_white_96}; 
although we find this does not raise the
predicted clustering to the observed values.
We thus discuss possibility (i) further below.
Nevertheless, it is quite remarkable that the predictions from our simple 
physical model fit the observed clustering over a 
wide range of magnitudes, scales and for the full range of $z=3-5$.

\subsection{Sensitivity to the halo density profile}
\begin{figure*}
\includegraphics[trim=0cm 6cm 0cm 0cm, clip=true, width =17.0cm, height=10.5cm, angle=0]
{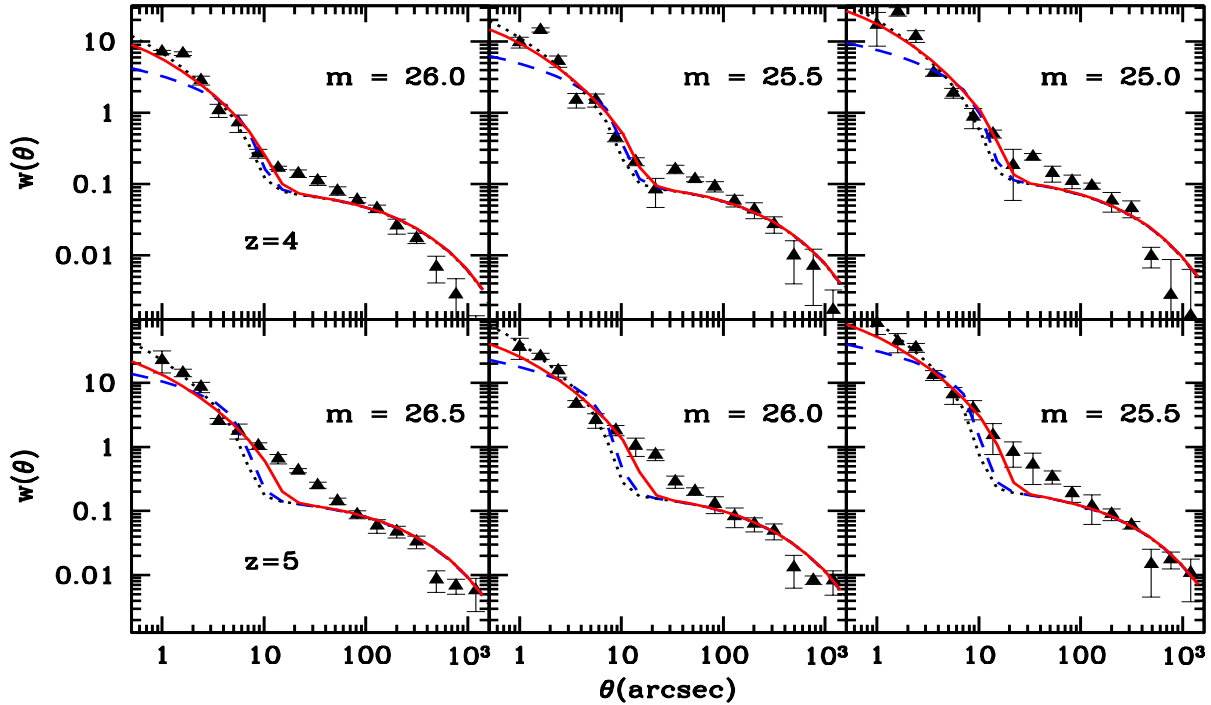} 
\caption{
The effect of concentration parameter and $s_{vir}$ on the clustering. 
Our fiducial model is shown in black dotted curve. The dashed blue 
curve is for a model where one uses concentration parameter given 
by \protect \cite{bullock_kolatt_01}. The solid red line shows our 
model predictions, that show the effect of extending the one halo term 
beyond the virial radius of the parent halo. 
In these cases, for simplicity, we multiplied the virial radius by factor $s_{vir}$. 
For $z=5$ we have used $s_{vir}=2$ and for $z=4$ we have used $s_{vir}=1.5$. 
Note that our fiducial model corresponds to $s_{vir} \sim 1$. 
The data points and error bars are from \protect \cite{hildebrandt_09_acf}. 
We used the fiducial cosmology parameters in this figure. 
All the astrophysical parameters the same as in Fig.~\ref{fig:acfz_full}. 
}
\label{fig:acf_models}
\end{figure*}

The discrepancy noted above, 
between the observed and our model predicted $w(\theta)$ at 
intermediate angular scales, 
could be due to one halo correlations of galaxies extending 
beyond the virial diameter of NFW density profile. 
Physically this 
means that subhalos (and satellite galaxies they host) 
can reside outside the virial diameter of the 
parent halo. If this is true, data suggest that, this effect is stronger 
for $z \geq 5$. 

To explore this possibility 
we assumed that halos follow a new density profile which is
of the same form as the NFW profile, 
but with a larger virial radius 
$\bar r_{vir} = s_{vir} r_{vir}$. We keep the 
the concentration parameter and the mass of the parent halo 
the same as before.
We took the fiducial values of $s_{vir} = 1.5$ and
$s_{vir} = 2.0$ at redshifts $4$ and $5$ respectively (see Table~\ref{tab3}).
Adopting $s_{vir} > 1$ can increase the clustering at intermediate scales but
will reduce the amplitude at small scales. This can be compensated by
adopting a smaller $\Delta t_0$.
In Fig.~\ref{fig:acf_models} we have shown the effect of 
extending the one halo term 
beyond the virial diameter of the parent halo. 
The value of $\Delta t_0$ is also reduced to $1.4 t_{dyn}$ and  
$1.3 t_{dyn}$ respectively at redshifts $4$ and $5$ 
to obtain a better fit at the smallest angular scales. 
The dotted black curve is for $s_{vir}=1.0$ (fiducial model). 
The curves obtained after extending the one halo 
term beyond the virial diameter of dark matter halos are shown in solid red lines. 
One can clearly see that these new curves provide a better fit to the observed 
clustering in small angular scales ($\theta<20''$). 
It is also clear from the figure that 
a change in the form of distribution of satellite 
galaxies inside the halos alone will not explain the discrepancy seen 
at $20'' \leq \theta \leq 80''$. 

We have also considered the prescription for the concentration parameter 
given by \citet{bullock_kolatt_01}, which is used
by earlier works high redshift clustering \citep{hamana_06}. 
The dashed blue curves in
Fig.~\ref{fig:acf_models} show the corresponding results
at $z=4$ and $5$.
One can clearly see that adopting the concentration parameter 
given by \citet{bullock_kolatt_01} underpredicts 
the clustering at $\theta \le 3''$

\begin{table}
\tabcolsep 5.2pt
\caption{
A summary of the fiducial parameters used in various models at at $z=4,5$ and $6$. 
The quantity $\kappa$ that determines the duration 
of star formation in our models is fixed to be $1$. 
For models where we extend one halo term beyond the virial diameter of a halo, 
we changed $\Delta t_0$ slightly from fiducial value. 
}
\label{table2}
\begin{tabular}{ccccccc}\hline
$z$ &$m$ &$\f{M_{agn}}{10^{12}M_\odot}$ &$\f{\Delta t_0}{t_{dyn}} $\T\B &$s_{vir}$ &$\chi_\nu^2(H)$ &$\chi_\nu^2(O)$\\
\hline 
~ \T  &24.5  &$0.8$    \T  &$1.6$     &$1.0$     &0.8 & \\   
3     &25.0  &$0.8$    \T  &$1.6$     &$1.0$     &0.4 &\\
~ \T  &25.5  &$0.8$    \T  &$1.6$     &$1.0$     &0.6 &\\

\hline

~ \T  &25.0  &$1.5$    \T  &$1.5$     &$1.0$     &3.6  &1.6 \\
4     &25.5  &$1.5$    \T  &$1.5$     &$1.0$     &1.4  &0.8 \\
~ \T  &26.0  &$1.5$    \T  &$1.5$     &$1.0$     &3.8  &2.9 \\
~ \T  &26.5  &$1.5$    \T  &$1.5$     &$1.0$     &     &1.3 \\
~ \T  &27.0  &$1.5$    \T  &$1.5$     &$1.0$     &     &0.2 \\
~ \T  &27.5  &$1.5$    \T  &$1.5$     &$1.0$     &     &0.3 \\
~ \T  &      &$1.5$    \T  &$1.4$     &$1.5$     &~    &~ \\

\hline

~\T   &25.5 &$3.0$    \T  &$1.4$     &$1.0$      &1.3  &\\
5     &26.0 &$3.0$    \T  &$1.4$     &$1.0$      &1.3  &\\
~ \T  &26.5 &$3.0$    \T  &$1.4$     &$1.0$      &5.7  &\\

~ \T  &     &$3.0$    \T  &$1.3$     &$2.0$      &      &\\

\hline
\end{tabular}
\label{tab3}
\end{table}

\subsection{Dependence on astrophysical and cosmological parameters}
In Section~\ref{sec:2halo_sensitivity}, 
we found that the large angular scale correlation functions
(with $80''< \theta < 1300''$) 
were fairly insensitive to
changes in the assumed astrophysical and cosmological parameters,
provided these models still fit the observed LF. We now ask
if this is true also for small angular scales below $\theta < 10''$
where our fiducial models can reasonably fit the observed data.
On small angular scales, the correlation functions can in 
principle be a function of astrophysical parameters as these change 
the total number of central and satellite galaxies satisfying the given 
luminosity criteria. 

\begin{figure*}
\includegraphics[trim=0cm 12cm 0cm 0cm, clip=true, width =17.0cm, height=7cm, angle=0]
{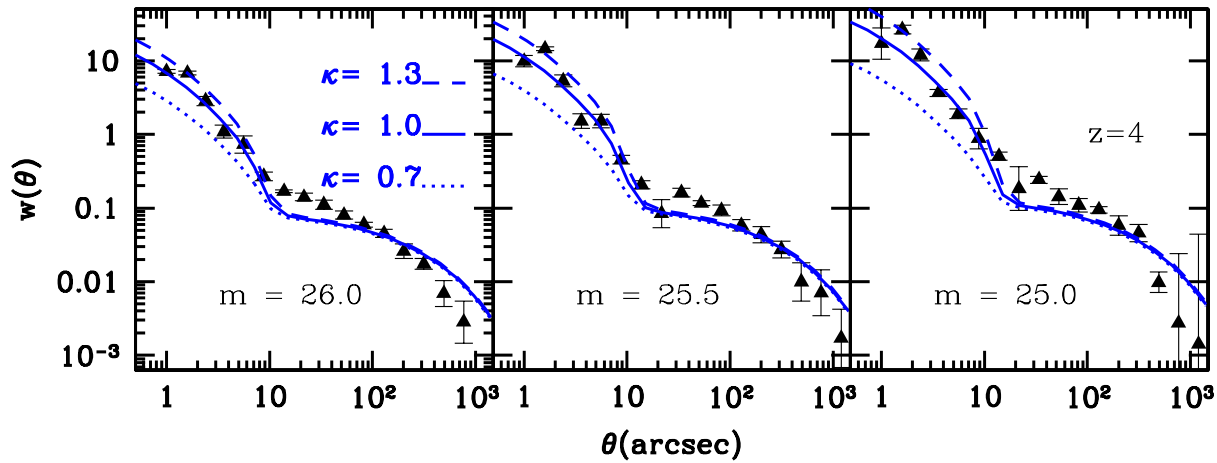} 
\caption{
The change in angular correlation function with $\kappa$  
around its fiducial value at $z=4$. The blue solid curve is the fiducial model 
prediction with $\kappa = 1.0$. The dotted and dashed curves are 
for $\kappa = 0.7$ and $1.3$ respectively. 
}
\label{fig:acf_z4_kappa}
\end{figure*}

In Fig.~\ref{fig:acf_z4_kappa}, we show the effect of varying 
$\kappa$ (which determines the duration of star formation activity in a halo) 
on small scale clustering. 
The blue solid curves are for our fiducial model with $\kappa = 1.0$. 
The blue dotted and dashed curves are for $\kappa$ values 0.7 and 1.3 
respectively. All other parameters except for $f_*/\eta$ have been kept
the same, and $f_*/\eta$ is varied such that the predicted LF still
fits the observed LF data reasonably well.
One can clearly see that clustering predictions especially 
for $\kappa = 0.7$ do not fit the correlation functions on small 
angular scales. 
Therefore we conclude that small scale clustering of LBGs at high redshifts 
can give useful information about $\kappa$ and thus 
the duration star formation in a halo.
We find that $\kappa \sim1$
or the star formation duration of order $t_{dyn}$ is favored
by the small scale clustering data.

We also varied $M_{agn}$, 
which determines the AGN feedback, 
around the fiducial value at each redshift,
ensuring that the predicted LFs still reasonably fits the and observed data.
We find that even the small scale angular correlation 
functions are fairly insensitive to these changes.
For example when we varied $M_{agn}$ at redshift $4$ from 
$1.2 \times 10^{12} M_\odot$ to $1.8 \times 10^{12} M_\odot$, 
the clustering predictions change at all scales 
by about $-7\%$ to $5\%$ from the fiducial model. 
We also found that the situation is similar for all magnitudes and 
other redshifts. 

We have also considered the effect of varying a whole suite of
cosmological parameters within their $2\sigma$ 
limits determined by WMAP7 year data. Again we ensure that the
predicted LFs best fits the observed data, by varying $f_*/\eta$.
We find that 
the small angular scale clustering is very insensitive 
to changes in all the cosmological parameters
except $\sigma_8$; where at most a $\pm20\%$ change results when 
$\sigma_8$ is varied between its $2\sigma$ limits  
given by WMAP 7 year data.

\section{Discussion and conclusions}

We have presented here a physically motivated semi-analytical 
model of galaxies to understand the clustering of high redshift 
Lyman break galaxies, where the model parameters are 
constrained by the observed high z luminosity function. 
For this purpose we use and expand upon 
the standard halo model.
Galaxies are assumed to be formed inside dark matter halos.
Their luminosity is determined by a physical model of star formation, 
which is a function of the mass and age of the hosting halo.

We began by assuming that each halo can host at most one visible galaxy.
On fitting the observed LF, we determine the relationship 
between the luminosity of a galaxy and the mass of its host halo. 
This allows us to calculate the large scale bias
for LBGs satisfying any luminosity threshold. This bias is
then folded in with the dark matter power spectrum to
predict the two point spatial correlation function
of these LBGs and hence the angular correlation function $w(\theta)$.
For our fiducial model, 
we find that the predicted $w(\theta)$ compares very well
with the observed data of 
both \citet{hildebrandt_09_acf} and \citet{ouchi_hamana_05_acf} 
at $\theta > 80''$, 
for the whole range of redshifts $z=3-5$ and 
limiting luminosities (magnitudes) 
(see Figs.~\ref{fig:acfz_lin},\ref{fig:acfz_full},
\ref{fig:acfz4_ouchi} and Table~\ref{tab3}). 

The predicted large scale galaxy bias $b_g$ agrees well with that
observationally determined by \citet{hildebrandt_09_acf} 
(see Table~\ref{tab1}).
At a given $z$ the bias increases with increasing luminosity, while
for a given $L_{th}$ it increases with $z$.
However, we find a smaller spread in $b_g$ as a function of 
$L_{th}$ at any given $z$ compared to that of \cite{hildebrandt_09_acf}. 
Remarkably, we find that the predicted large scale clustering of LBGs
is fairly insensitive to the assumed astrophysical or cosmological
parameters, provided we simultaneously fit the observed LF. 
This may point to an important internal consistency of
our physical model; that if we fix the mass to light ratio
correctly by using the LFs of LBGs, then the standard LCDM model
correctly predicts the amplitude of their large scale clustering.

We then extended our approach by incorporating the
Halo occupation distribution, which provides
the distribution of galaxies inside dark matter halos.
This is separated into the central, $f_{cen}$, 
and satellite, $N_s$, contributions.
Often the HOD is modelled in a parametrized form. Instead we have
adopted a more physical approach.  
We use our prescription for computing the LF to estimate 
$f_{cen}$. The conditional mass function and our star formation 
model are used to calculate the mean number 
and luminosity of satellite galaxies in a parent halo.
An additional parameter, $\Delta t_0$, is introduced in calculating $N_s$
(see Table~\ref{tab0}). If parent halo collapses
at a time $t_p$, a subhalo can host a detectable galaxy only 
if it collapses at an earlier epoch, $t_s < t_p - \Delta t_0$.
In order to explain the small angular scale
clustering, one requires $\Delta t_0$ of order $t_{dyn}$, 
the dynamical time scale of the parent halo. A much smaller (or
larger) $\Delta t_0$ leads generically to an excess (deficit) of 
small angular scale clustering.

The calculated forms of $f_{cen}(M,L_{th},z)$ and $N_s(M,L_{th},z)$
compare reasonably with that assumed in parametrized models 
of \citet{hamana_06} and the simulations of \citet{conroy_wechsler_06}.
The average value of $f_{cen}$ is typically $0.4$. Thus,
on the average, $40\%$ of the halos above a minimum mass $M_{min}$
(which itself depends on the luminosity threshold), at any given redshift,
can host detectable central galaxies.
Further the average value of $N_s$ is about $0.02-0.04$, or about $5-10\%$
of the halos with detectable central galaxies, also will have a
detectable satellite. Indeed it is such pairs which contribute
to the small angular scale clustering.
At $z = 4$ and for apparent magnitude thresholds in the range 25-26,
the average mass of halos contributing to the observed clustering, $M_{av}$,
ranges from $10^{12} M_\odot -3.9 \times  10^{11} M_\odot$. 
At $z = 5$ these masses are smaller by a factor $\sim 1.4$, 
while at $z = 3$ these masses are 
larger by a similar factor. At any given redshift and magnitude
threshold, the typical mass $M_P$ of parent halos hosting
detectable satellite galaxies, are about 2 times larger than $M_{av}$. 

Having obtained the HOD, 
we can calculate both the one halo and two halo 
contributions to the total correlation function of LBGs.
Our simple physical model gives a reasonable 
fit to the observed clustering of LBGs 
at all angular scales for the faintest LBGs 
with $m \ge 25$ at $z=3$ and for $m \ge 26.5$ at $z=4$.  
At $z=5$, and for the most
luminous galaxies at $z=3,4$, 
the predicted $w(\theta)$
again fits the observed data well
at both large ($\theta>80''$) and small ($\theta < 10''$) angular scales.
The clustering at small angular scales 
as mentioned above, is likely to be dominated by 
pairs of LBGs rather than rich clusters, as
the number of detectable galaxies hosted by
most collapsed halos is typically less than 2.

The small angular scale clustering is also not very sensitive
to changes of several cosmological and astrophysical parameters
from their fiducial values, as long as we simultaneously fit 
the observed LF. 
However, the amplitude of $w(\theta)$ on
small angular scales is very sensitive to the 
value of $\kappa$, which determines the 
duration of star formation activity in a halo.
The present data are consistent 
with $\kappa \sim 1$ or a star formation duration of the order of the dynamical 
time scale $t_{dyn}$ of the dark matter halo.

We find that the following broad physical picture of LBGs 
consistently accounts for their observed LF and clustering.
First the average mass of the halos hosting the brightest 
central LBGs at $z=3-5$, with $-21< M_{AB} < -20$, 
is around $3\times 10^{11} M_\odot$ to $1.5 \times 10^{12} M_\odot$.
Halos which host detectable
satellites and contribute dominantly to the small angular 
clustering are more massive by a factor of 2 or so.
Typically fainter LBGs or those at higher $z$ 
are hosted by smaller mass halos.
In these galaxies about $50\%$ of the stars are formed 
over a timescale of $300-500$ Myr for $z=5-3$,
by converting $\sim 3-8\%$ of the baryons to stars. 
Our physical model for the HOD suggests that
approximately $40\%$ of the halos above a minimum mass $M_{min}$,
can host detectable central galaxies.
This is comparable to the
duty cycle values preferred by \citet{lee_09,lee_ferguson_12}.
Further, about 
$5-10\%$ of these halos are likely to also host a detectable satellite.
These satellites form over a dynamical timescale or so prior to
the formation of the parent halo.
The small angular scale clustering is mainly due to central-satellite 
pairs. The average fraction of halos which can host a central LBG can be
compared to the duty cycle invoked in the literature.
Finally, a preliminary study suggests that
the star formation model that we have invoked
is also consistent with observation of the SFR-$M_*$ relation,
and the stellar mass function.

The theoretically predicted $w(\theta)$ at 
intermediate angular scales is smaller than that observed,
for the brightest LBGs at $z=3,4$ and at $z=5$.
We explored in detail whether this excess can be due to
a more extended satellite galaxy distribution.
This only partly accounts for the discrepancy.
We also find that the  non-linear corrections to 
dark matter power spectrum does not raise the predicted 
clustering at the intermediate scales to the observed values. 
Note that the typical mass of dark matter halo hosting a galaxy increases
with its luminosity. Also the higher order (quasi-linear) corrections to 
the dark matter halo bias 
is expected to be larger for higher mass and higher redshift halos,
which are rarer. Therefore we   
suspect that the higher order quasi-linear corrections to galaxy bias 
could be playing a role in explaining the excess intermediate
scale clustering, 
a possibility which we hope to
explore in the future.
Nevertheless, it is noteworthy that the predictions 
from our simple physical model, employing only a few
free parameters, can fit the observed clustering data over 
a wide range of scales, redshifts and limiting luminosities. 

\section*{Acknowledgments}
We thank Hendrik Hildebrandt, Masami Ouchi and Nobunari Kashikawa 
for providing the 
observed data of angular correlation functions.
CJ thanks Aseem Paranjpaye for many discussions while writing the
code to incorporate the halo model. CJ also acknowledges support
from CSIR. KS acknowledges partial support from NSF grant 
PHY - 0903797 while visiting University of Rochester and Eric Blackman at 
Rochester for warm hospitality. 
SS acknowledges support from University of Kwazulu-Natal, South Africa. 
We thank an anonymous referee for very detailed and useful
comments which helped us to improve our paper.

\def\aj{AJ}%
\def\actaa{Acta Astron.}%
\def\araa{ARA\&A}%
\def\apj{ApJ}%
\def\apjl{ApJ}%
\def\apjs{ApJS}%
\def\ao{Appl.~Opt.}%
\def\apss{Ap\&SS}%
\def\aap{A\&A}% 
\def\aapr{A\&A~Rev.}%
\def\aaps{A\&AS}%
\def\azh{AZh}%
\def\baas{BAAS}%
\def\bac{Bull. astr. Inst. Czechosl.}%
\def\caa{Chinese Astron. Astrophys.}%
\def\cjaa{Chinese J. Astron. Astrophys.}%
\def\icarus{Icarus}%
\def\jcap{J. Cosmology Astropart. Phys.}%
\def\jrasc{JRASC}%
\def\mnras{MNRAS}%
\def\memras{MmRAS}%
\def\na{New A}%
\def\nar{New A Rev.}%
\def\pasa{PASA}%
\def\pra{Phys.~Rev.~A}%
\def\prb{Phys.~Rev.~B}%
\def\prc{Phys.~Rev.~C}%
\def\prd{Phys.~Rev.~D}%
\def\pre{Phys.~Rev.~E}%
\def\prl{Phys.~Rev.~Lett.}%
\def\pasp{PASP}%
\def\pasj{PASJ}%
\def\qjras{QJRAS}%2215.bib
\def\rmxaa{Rev. Mexicana Astron. Astrofis.}%
\def\skytel{S\&T}%
\def\solphys{Sol.~Phys.}%
\def\sovast{Soviet~Ast.}%
\def\ssr{Space~Sci.~Rev.}%
\def\zap{ZAp}%
\def\nat{Nature}%
\def\iaucirc{IAU~Circ.}%
\def\aplett{Astrophys.~Lett.}%
\def\apspr{Astrophys.~Space~Phys.~Res.}%
\def\bain{Bull.~Astron.~Inst.~Netherlands}%
\def\fcp{Fund.~Cosmic~Phys.}%
\def\gca{Geochim.~Cosmochim.~Acta}%
\def\grl{Geophys.~Res.~Lett.}%
\def\jcp{J.~Chem.~Phys.}%
\def\jgr{J.~Geophys.~Res.}%
\def\jqsrt{J.~Quant.~Spec.~Radiat.~Transf.}%
\def\memsai{Mem.~Soc.~Astron.~Italiana}%
\def\nphysa{Nucl.~Phys.~A}%
\def\physrep{Phys.~Rep.}%
\def\physscr{Phys.~Scr}%
\def\planss{Planet.~Space~Sci.}%
\def\procspie{Proc.~SPIED}%
\let\astap=\aap
\let\apjlett=\apjl
\let\apjsupp=\apjs
\let\applopt=\ao

\bibliographystyle{mn2e}	% (uses file "plain.bst")
\bibliography{ref.bib,sfr.bib}		% expects file "myrefs.bib"

\appendix
\section{Clustering dependence on $\sigma_8$}
For any mass scale $M$, the variance of smoothed density contrast 
$\sigma^2(M) \propto ~ \sigma^2_8 k^3_M P(K_M) \sim \sigma^2_8 k_M^{3+n_{eff}}$. 
Here $n_{eff}$ is the effective spectral index, which is $\sim -2$ on galatic 
scales and $-1$  on cluster scales. We also have $k_M^{-1} \sim M^{1/3}$. Thus we get 
\be
\sigma^2(M) \propto \sigma^2_8~ M^{\f{-(3+n_{eff})}{3}}.
\label{eqn:A1}
\ee
For high redshifts $\nu(M,z) = (\dl_c(z)/\sigma(M))^2 >> 1$.
Hence using Eq.~\ref{eqn:halo_bias} we get the halo bias, 
$b(M,z) \propto \nu(M,z) \propto 1/\sigma^2(M)$ 
Using \ref{eqn:A1} in this, the scaling of halo bias at high redshifts can 
be written as
\be
b(M,z) \propto \df{M^{\f{3+n_{eff}}{3}} }{\sigma^2_8}.
\label{eqn:A2}
\ee
We know that on large scales galaxy bias $b_g(k,z,L_{th})$ 
given by Eq.~(\ref{eqn:biasL2}) is a constant (or k independent). 
To get the scaling behavior of this galaxy bias we assume that in 
Eq.~(\ref{eqn:biasL2}) the major contribution to galaxy bias comes from masses  
at and around $M_{av}$. In this limit $b_g(k,z,L_{th}) \sim b(k,M_{av})$. 
The quantity $M_{av}$ is the average mass of the halo that can host a galaxy of 
luminosity $L_{th}$ (See Fig.~\ref{fig:zc_mass}). 
It is clear that $M_{av}$ depends on astrophysical and cosmological 
parameters. Also, as discussed in Section~\ref{sec:xi_linear}, on very 
large scales galaxy bias is scale independent. 
Thus using Eq.~(\ref{eqn:A2}) we get 
\be
b_g(z,L_{th}) \propto \df{M_{av}^{\f{3+n_{eff}}{3}} }{\sigma^2_8}.
\label{eqn:A3}
\ee
The large scale (small $k$) dark matter power spectrum $P_{lin}(k)$ always scale as 
square of $\sigma_8$. Hence using Eq.~(\ref{eqn:pgl}) and Eq.~(\ref{eqn:A3}) we get 
the follwing scaling law of galaxy power spectrum. 
\be
P^{2h}_g(k,L_{th}, z) = b^2_g(L_{th},z) P_{lin}(k,z)
       \propto ~\df{ M_{av}^{\f{2(3+n_{eff})}{3}} }{\sigma^2_8} \hat F(k,z)
\ee
where  $P_{lin}(k,z)= \sigma_8^2~\hat F(k,z)$.
The two point correlation functions on large scales also scale in 
this way. Thus under the above approximations
\be
\xi_{2h}(L_{th}, R,z) \propto ~ \df{ M_{av}^{\f{2(3+n_{eff})}{3}} }{\sigma^2_8} F(R,z)
\ee
where $F(R,z)$ is the transformation of $\hat F(k,z)$ as given in 
Eq.~(\ref{eqn:xir-2h}) 
Suppose one assumes two models A and B with 
different values of $\sigma_8$, say $\sigma^A_8$ and $\sigma^B_8$. 
The ratio between two point correlation functions of both models 
on large scales will be
\be
\f{\xi^A_{2h}}{\xi^B_{2h}} = \left(\f{\sigma^B_8}{\sigma^A_8}\right)^2 
             \left(\f{M^A_{av}}{M^B_{av}}\right)^{\f{2(3+n_{eff})}{3}} 
\label{eqn:A5}
\ee
For galaxies we have $n_{eff} \sim -2$. 
Hence the ratio between galaxy two point correlation 
functions of models A and B is given by 
\be
\f{\xi^A_{2h}}{\xi^B_{2h}} = \left(\f{\sigma^B_8}{\sigma^A_8}\right)^2 
             \left(\f{M^A_{av}}{M^B_{av}}\right)^{\f{2}{3}} 
\label{eqn:A6}
\ee

\end{document}